\documentclass{sig-alternate}
\usepackage{framed}
\usepackage{times}
\usepackage{color}
\usepackage{url}
\usepackage{graphicx}
\begin{document}
\newtheorem{lemma}{Lemma}[section]
%
% --- Author Metadata here ---
%\conferenceinfo{WOODSTOCK}{'97 El Paso, Texas USA}
%\CopyrightYear{2007} % Allows default copyright year (20XX) to be over-ridden - IF NEED BE.
%\crdata{0-12345-67-8/90/01}  % Allows default copyright data (0-89791-88-6/97/05) to be over-ridden - IF NEED BE.
% --- End of Author Metadata ---

%\title{A New Structure Index for XML}
%\title{TinyTree: A Fast Structural Self-Index for XML}

\title{Fast and Tiny Structural Self-Indexes for XML}

%\titlenote{A full version of this paper is available as
%\textit{Author's Guide to Preparing ACM SIG Proceedings Using
%\LaTeX$2_\epsilon$\ and BibTeX} at
%\texttt{www.acm.org/eaddress.htm}
%}
\numberofauthors{2} 
\author{
\alignauthor
Sebastian Maneth\\
       \affaddr{NICTA and UNSW}\\
       \affaddr{Sydney, Australia}\\
       \email{sebastian.maneth@nicta.com.au}
% 2nd. author
\alignauthor
Tom Sebastian\\
       \affaddr{NICTA}\\
       \affaddr{Sydney, Australia}\\
       \email{tom.sebastian@nicta.com.au}
}
\additionalauthors{Additional authors: John Smith (The Th{\o}rv{\"a}ld Group,
email: {\texttt{jsmith@affiliation.org}}) and Julius P.~Kumquat
(The Kumquat Consortium, email: {\texttt{jpkumquat@consortium.net}}).}
\date{30 July 1999}
\maketitle
\begin{abstract}
XML document markup is highly repetitive and therefore well
compressible using dictionary-based methods such as DAGs or
grammars. In the context of selectivity estimation,
grammar-compressed trees were used before as synopsis for
structural XPath queries. 
% can be evaluated over grammar-compressed trees. 
Here a fully-fledged index over such grammars
is presented. The index allows to execute arbitrary tree algorithms
with a slow-down that is comparable to the space improvement.
More interestingly, certain algorithms execute much faster
over the index (because no decompression occurs). E.g., for structural
XPath count queries, evaluating over the index is faster than  
previous XPath implementations, often by two orders of magnitude.
The index also allows to serialize XML results (including texts)
faster than previous systems, by a factor of ca. $2$--$3$. This is due
to efficient copy handling of grammar repetitions, and because
materialization is totally avoided. In order to compare with
twig join implementations, we implemented a materializer 
which writes out pre-order numbers of result nodes, and
show its competitiveness.
%that it is comptetitive.
\end{abstract}

\section{Introduction}
An important task in XML processing is the evaluation
of XPath queries. Such queries select nodes of an
XML document and are used in many scenarios: embedded in
larger XQueries, in XSL stylesheets, in XML policy
specifications, in JavaScripts, etc.
A common way of speeding up query evaluation is to use
indexes. But conventional value indexes for XML tags
and text values are not sufficient to answer
XPath queries, because they do not capture the document's 
hierarchical structure. Therefore a large number
of structural XML indexes have been introduced
(see~\cite{DBLP:journals/is/FletcherGWGBP09} for a
recent overview).
The first one was the DataGuide~\cite{DBLP:conf/vldb/GoldmanW97}.
It stores a summary of 
all distinct paths of a document. Later the finer 1-index
was proposed~\cite{DBLP:conf/icdt/MiloS99} which is based on
node bisimulation. 
For certain XPath queries these indexes allow evaluation
without accessing the original data; e.g., 
for structural queries restricted to the child and descendant axes.
More fine-grained structural indexes were considered
but turned out to be too large in practice,
see~\cite{DBLP:conf/sigmod/KaushikBNK02}. 
As a compromise, the A($k$)-index~\cite{DBLP:conf/icde/KaushikSBG02}
was proposed which uses node bisimilarity of paths up to length $k$;
the D($k)$~\cite{DBLP:conf/sigmod/QunLO03} and
M($k$)-indexes~\cite{DBLP:conf/icde/HeY04} 
are A($k$)-variants that adapt 
to query workloads. Updates for the A($k$) and 1-indexes
were studied in~\cite{DBLP:conf/sigmod/YiHSY04}. Index path selection
is considered 
e.g., in~\cite{DBLP:conf/xsym/RunapongsaPBP04}; but, their 
indexes are usually larger than the original
documents (including data values).
All indexes mentioned so far are approximative for full
structural XPath, i.e., do not capture enough
information to evaluate XPath's twelve navigational axes. 
This is in contrast to the indexes introduced here.

A \emph{self index} has the property that
(1) it allows to speed up certain accesses, and 
(2) it can reproduce the original data 
(which therefore can be discarded after index construction).
Moreover, such indexes are often based on compression and
hence are \emph{small} (typically smaller than the original data). 
For Claude and Navarro~\cite{clanav10} a self-index  for text
must (at least) efficiently support the extract and find operations;
these operations reproduce a portion of the text and find all occurrences
of a substring, respectively.
In XPath processing, more complex search than finding 
substrings is required. 
In fact, XPath search is comparable to regular expression search.
Unfortunately, even for text, little is known about 
indexes that support arbitrary regular expression search (see,
e.g.,~\cite{DBLP:journals/jacm/Baeza-YatesG96}). 

In~\cite{arr+10,DBLP:journals/pvldb/ManethN10} it was
observed that two particular navigational operations allow drastic
speed-ups for XPath evaluation: taggedDesc and taggedFoll. Given a
node 
and a label, these operations return the first descendant node
and first following node with that label, respectively.
During XPath evaluation these operations 
allow to \emph{jump} to next relevant nodes;
this cuts down the number of intermediate nodes
to be considered during evaluation.
The ``QName thread'' in
MTree~\cite{DBLP:conf/sac/PettovelloF06} is similar in spirit
(it allows to jump to next descendants with a given
label).

The idea of our new index is to use grammar-compressed trees
(which typically are much smaller than succinct
trees~\cite{DBLP:conf/dexaw/ManethMS08}) and to add small data
structures 
on top of these which support efficient taggedDesc and taggedFoll.
Our contributions are 
\begin{enumerate}
\item a self-index for trees, based on grammar-based compression
\item a generic sequential interface that allows to execute,
over the new index, arbitrary algorithms on the original tree
\item a special evaluator for \emph{counting} of XPath query results, and
\item special evaluators for \emph{serializing} and
\emph{materializing} of XPath query results.
\end{enumerate}
We tested the generic interface of Point~2 on two algorithms:
on depth-first left-to-right (dflr) recursive and iterative 
full tree traversals,
and, on the (recursive) XPath evaluator ``SXSI'' of~\cite{arr+10}.
We obtain good time/space trade-offs.
For instance, replacing SXSI's tree store with our interface of
Point~2 gives a slow-down of factor $4$ while it slashes SXSI's memory use
by factor $3$ (averaged over the $16$ tree queries of~\cite{arr+10} on
a 116M XMark file).
Our experiments show that the
evaluators of Points 3 and 4 are faster than existing XPath 
implementations, often by two orders of magnitude. 
Note that the indexes used by these evaluators
are so \emph{tiny} in space (see Figure~\ref{fig:TotIndexSizes})
that any XML database can profit from them, by
conveniently keeping them in memory. This allows, besides
others, fast serialization and fast XPath selectivity computation,
and therefore can replace structural synopses. 
\begin{figure}[ht!]
\begin{center}
{
\small
\begin{tabular}{|l|c|c|c|c|c|}
\hline
&\multicolumn{3}{c|}{XMark}&\multicolumn{1}{c|}{Treebank}&
\multicolumn{1}{c|}{Sprot}\\
\cline{2-6}
 & 116MB & 1GB & 11GB & 83MB & 437MB \\
\hline
Count & 1.1 & 5.5 & 7.9 & 3.5 & 1.3 \\
Mat. / Ser. & 1.7 & 1.9 & 11.6 & 5.5 & 2.5 \\
\hline
\end{tabular}
\caption{TinyT Index Sizes (in MB)}
\label{fig:TotIndexSizes}
}
\end{center}
\end{figure}

While the generic interface causes a slow-down due to
decompression, there are classes of algorithms (over
the grammar) which allow considerable \emph{speed-ups}.
Essentially, the
speed-ups are proportional to the compression ratio, because
the compressed grammar need only be traversed once.
For instance, tree automata and Core XPath can be evaluated
in one pass over 
straight-line tree (SLT) grammars~\cite{lohman06}.
This idea was used in~\cite{fisman07} for selectivity
estimation of structural XPath. They study 
synopsis size and accuracy, but do not consider
efficient evaluation.
We combine the ideas of~\cite{DBLP:journals/pvldb/ManethN10,arr+10}
with that  
of evaluating in one pass over the grammar. 
To this end, we augment the grammar with information that allows
efficient taggedDesc and taggedFoll: for every nonterminal
$X$ and terminal symbol $t$ of the grammar a bit is stored that
determines  whether $X$ generates $t$.
If $X$ does \emph{not} generate $t$, then it may be 
``jumped'' during a taggedDesc call for $t$.
Our first structural index comprises 
this ``jump table'', together with a compact representation
of the grammar.
The XPath count evaluator of Point~3 executes over
this index. 
To obtain grammars from XML structure trees, we use
the new TreeRePair compressor~\cite{lohmanmen10}.
Due to compression, the resulting indexes
are phenomenally small. For instance (cf. Figure~\ref{fig:TotIndexSizes}), 
our index can store the half a billion nodes of an 11GB XMark tree
in only 8MB!
This means an astonishing 8.7 nodes per bit!
Consequently, our evaluator (which, due to jumping,
need not even visit the whole grammar)
is extremely fast.
Compared to the fastest known evaluators,
MonetDB~\cite{DBLP:conf/sigmod/BonczGKMRT06} and Qizx~\cite{qizx}, 
we found that our XPath count evaluator
is faster by 1--2 orders of magnitude, 
for essentially all queries we tested.

Motivated by our positive results from Point~3,
the question arose whether the count
evaluator can be extended to handle proper XPath semantics, i.e., 
to output XML subtrees of result nodes.
Since serialization involves outputting of data values, 
all data values are now stored in a memory buffer.
Additionally, a data 
structure is built that links the SLT grammar to the correct data values.
This is achieved by storing for each nonterminal the number of
text-values that it generates. In fact, since a nonterminal generates
a \emph{tree pattern} which has many ``dangling subtrees'',
we need to store tuples of such numbers: the first component is the
number of text-values in the first ``chunk'' of the nonterminal,
i.e., in the tag sequence (of the nonterminal)
before the first dangling subtree; next
is the number of text-values in the second chunk,
i.e., between the first and second dangling subtrees, etc.
Evaluation is still done in one pass through the grammar, but,
this time  must follow a strict dflr traversal 
(which makes it slower than for count queries). 
A nice bonus is the possibility to 
make clever use of hashing: we remember the ``chunks'' of
XML markup produced by each nonterminal. This greatly speeds
up serialization. Moreover, it turned out that
materialization of result nodes can be totally avoided.
Thus, neither expensive grammar node IDs need to be stored, 
nor their translation to pre-order numbers is needed.
Rather, whenever a result node is encountered during evaluation,
we start a serialization process 
which works in parallel with evaluation. 
The resulting system outperforms by a factor of $2$--$3$ 
the fastest known system SXSI (which on its own outperforms 
MonetDB and Qizx, see~\cite{arr+10}). 

About the comparison: it can be argued that 
comparing our rudimentary XPath evaluator with
full-blown XML databases is unfair, because these larger systems 
have more overhead (such as locking, transaction handling, updates).
On the other hand, these systems are highly optimized and therefore
could exploit their best available algorithm for simple queries.
We therefore believe that the comparison is relevant.
Note that we also compare with specialized implementations
which handle smaller or incomparable XPath fragments.
For instance, we compared to the fastest available implementations
of twig 
joins~\cite{DBLP:journals/pvldb/GrimsmoBH10,DBLP:conf/dbkda/GrimsmoBT10}.
Since these algorithms
materialize result nodes, we  
implemented an experimental materializer (Point~4).
Interestingly, it often outperforms these twig implementations
(which represent state-of-the art of many years
of research on twigs).
We also compare to the index 
of~\cite{ferlucmanmut05,DBLP:conf/www/FerraginaLMM06} which
handles simple paths (XPaths with one // followed by /'s); 
our experiments show that for selective queries this index is faster than ours
(by a factor of $10$--$20$), while for non-selective queries
our index is faster.

%\subsubsection*{Related Work}
\paragraph*{Related Work}

Compression by SLT grammars was used in~\cite{fisman07} for
selectivity estimation of structural XPath. They study the
space efficiency of binary encoded grammars with respect
to other XML synopses, but do not study run times.
It is also shows that updates can be handled incrementally
with little space overhead; this is important also for our work,
because we would like to support incremental updates in the future.
The minimal DAGs used by Koch et al.~\cite{bungrokoc03_short} can be
seen as the first 
grammar-compressed approach to XML trees (a DAG naturally 
corresponds to a regular tree grammar). For usual XML document trees,
minimal DAGs only exhibit $10\%$ of the original number of edges.
More powerful grammar-compressors such as BPLEX~\cite{buslohman08}
further reduce this number to $5\%$ and the recently introduced 
TreeRePair~\cite{lohmanmen10} to less than $3\%$.
An SLT grammar generalizes DAGs from sharing of repeated subtrees 
to \emph{sharing of repeated tree patterns} (connected subgraphs 
of the tree). They are equivalent to the sharing graphs
used by Lamping for optimal lambda calculus evaluation~\cite{lam90}.
A self-index for grammar-compressed strings was presented
in~\cite{clanav10}. They show efficient support for
extract and find. 
It can be shown, but goes beyond the scope of this paper,
that the extract operation can be generalized from
their string grammars to our SLT grammar, with the same
time bounds as in their result.
In~\cite{arr+10} they use the succinct tree data structures
of~\cite{sadnav10} and add explicit copies for each label,
using compressed bit-arrays~\cite{okasad07}. This allows constant time
access to taggedDesc and taggedFoll (using rank and select over
bit-arrays), but becomes fairly memory heavy 
(for a 116M XMark document with 6 million nodes, they need
8MB for the tree, and additional 18MB to
support taggedDesc and taggedFoll in constant time). 
%Note that our index, in principle, fits with the general framework
%of~\cite{DBLP:conf/sigmod/KaushikKNR04} where a structural index
%is defined as a path-preserving ``contraction'' of the structure tree.
%also mention the general framework of ``integrated use of
%structural and value indexes''.

\section{XML Tree Compression}

An XML document naturally corresponds to an unranked ordered
tree. For simplicity, we only focus on element nodes, attributes,
and text values, and omit namespaces, processing instructions,
and comments.
Our data model assumes that the attribute and text values are
stored separately from the tree structure (in a ``text collection''), 
and that they can be addressed by a function getText($n$) that 
returns the $n$-th text or attribute value (in pre-order appearance). 
In our tree model, text nodes of the document are
represented by placeholder leaf nodes labeled
by the special label \_T.
Similarly, attribute definitions are represented by 
``attribute placeholder nodes'' labeled \_A; such a
node has children nodes which are labeled by the names of the attributes
(prepended by the symbol ``@'') in their appearance order,
which themselves have a single ``attribute-text placeholder
node'' (labeled \_AT).
For instance the XML element
$<$name id="9" r="4"$>$Text$<$/name$>$
is represented, in term syntax, by this tree:
name(\_A(@id(\_AT),@r(\_AT)),\_T).
For a given XML document, such a tree is called 
its \emph{XML structure tree}.
\begin{figure}[ht!]
{\small
\centerline{
\begin{tabular}{|l|*{3}{@{\hspace{5pt}}c@{\hspace{5pt}}|}}
\hline
Name & Element Count & Max Depth & Non-Text (MB)\\
\hline
XMark116M   & 1,735,083 & 12 & 34\\
XMark1G & 16,703,210 & 13 & 325\\
XMark11G & 167,095,844 & 13 & 3246\\
%\hline
Treebank83M & 2,437,666 & 22 & 25\\
%\hline
Sprot437M & 10,903,568 & 38 & 154\\
\hline
\end{tabular}}}
\caption{Datasets used in experiments}
\label{fig:docs}
\end{figure}
\begin{figure}[ht!]
{\small
\centerline{
\begin{tabular}{|l|*{2}{@{\hspace{5pt}}c@{\hspace{5pt}}|}}
\hline
Name & Element Count & Size (MB)\\
\hline
XMark116M & 6,074,297 & 59\\
XMark1G & 58,472,941 & 559\\
XMark11G & 584,961,650 & 5579\\
%\hline
Treebank83M & 7,312,615 & 48\\
%\hline
Sprot437M & 27,035,515 & 231\\
\hline
\end{tabular}
}}
\caption{Sizes of XML structure trees}
\label{fig:structureTree}
\end{figure}
Obviously, these trees are larger than pure element-node
trees, because of the additional placeholder nodes.
The placeholder nodes help to get from a node in the structure tree to
the corresponding value (by keeping track of how many placeholders
have appeared so far). Moreover, they allow to answer certain
queries directly on the structure index, such as, e.g., the
query //text().
%This tree model is similar to the one of~\cite{arr+10}, only
%that they work over unranked trees (represented as balanced
%parenthesis structures).
To get a rough estimate of the different node counts for
element only trees and their corresponding XML structure trees,
see Figures~\ref{fig:docs} and~\ref{fig:structureTree}.
The ``Non-Text'' numbers refer to the sizes of XML files in which
all text and attribute values where cut out (thus yielding
non-valid XML). If those values are replaced by our placeholder
nodes, then we obtain the XML structure trees whose sizes are shown in
Figure~\ref{fig:structureTree} (their depth changes 
at most by two, due to attribute placeholders).
The XMark files were generated with the XMark generator
(\url{http://www.xml-benchmark.org}),
Sprot437M is the protein databased used in~\cite{bungrokoc03_short},
and Treebank83M is a linguistic database obtained
from \url{http://www.cs.washington/edu/research/xmldatasets}.

In our model, an XML structure tree is represented by a
\emph{binary tree} which stores the
first-child and next-sibling relationship of the XML document
in its first and second child, respectively.
%If a node had no first-child or next-sibling, then 
%a null node (labeled by the special label ``\_'') is
%inserted in our binary tree. Thus, our binary trees have
%roughly twice the number of nodes than the XML structure trees,
%if  null-nodes are counted. This is an unimportant
%technicality, because our compression factors out these 
%repeated null nodes.
%We therefore do not mention them any further.
The idea of grammar-based tree compression is to find a small
tree grammar that represents the given tree.
For instance, the minimal unique DAG of a tree can be
obtained in amortized linear time (see,
e.g.,~\cite{bungrokoc03_short}); it can be seen as a particular
tree grammar (namely, a regular one).
For instance, the minimal DAG for the binary tree
$t = f(f(a(b,c),a(c,c)),f(c,c))$ can be written as this grammar:
\[
\begin{array}{lcl}
S&\to& f(f(a(b,C),A),A)\\
A&\to& a(B,B)\\
B&\to& c
\end{array}
\]
The \emph{size} of a grammar is the total number of edges of the trees
in the right-hand sides of its productions.
The grammar in the above example has size $8$.
In contrast, the original tree has size $10$.
In our grammars there is exactly one production for each
nonterminal $A$. The \emph{right-hand side} of $A$'s production
is denoted by $\text{rhs}(A)$.
We fix $\sigma$ as the size of the alphabet of a grammar,
consisting of terminal and nonterminal symbols.

In an SLT grammar, sharing is not restricted to
subtrees, but arbitrary \emph{tree patterns}
(connected subgraphs) can be shared.
In the example tree $t$ of above,
the tree pattern consisting of an $f$-node and right subtree
$a(c,c)$ appears twice. As we can see, this tree pattern
has one ``dangling edge'', namely, to the second-child of
the $f$-node.
In SLT grammar notation, a tree pattern is written as a tree
in which special placeholders, called \emph{parameters},
are inserted at dangling edge positions. The parameters
are denoted $y_1, y_2, \dots$ and are numbered in the order
of appearance of dangling edges.
An SLT grammar that represents $t$ has these productions:
\[
\begin{array}{lcl}
S&\to& A(A(a(b, c)))\\
A(y_1)&\to& f(y_1, a(c, c))
\end{array}
\]
The nonterminal $A$ uses one parameter $y_1$ to represent the single
dangling edge of the pattern mentioned above. The 
size of this grammar is still $8$.
The number of parameters of a nonterminal $A$ is called
its \emph{rank} and is denoted $\mbox{rank}(A)$. 
The maximal number of parameters of the nonterminals of a 
grammar is called the \emph{rank} of the grammar.
Another important aspect of a grammar is its \emph{depth}, which
is the length of the 
longest sequence of nonterminals $A_1,A_2,\dots, A_d$ 
such that $A_{i+1}$ appears in the right-hand side of $A_i$,
for all $1\leq i< d$.  
Since all our grammars produce one tree only, 
the depth is bounded by the number of
nonterminals. Given a nonterminal $A$ (of rank $k$),
its \emph{pattern tree}, denoted $t_A$, is the tree
over terminal symbols and parameters $y_1,\dots,y_k$, obtained
from $A(y_1,\dots, y_k)$ by applying grammar productions
(until no production can be applied anymore).

While the minimal DAG of a tree is unique and can be found 
in linear time, the minimal SLT grammar is not unique, and
finding one is NP-complete~\cite{buslohman08}.
%(this already holds for
%ordinary cf grammars on strings~\cite{cha+05}). 
The BPLEX approximation algorithm~\cite{buslohman08}
generates SLT grammars that are ca. half the size of the minimal DAG.
The new TreeRePair algorithm~\cite{lohmanmen10} improves this by
another $20\%$--$30\%$ (while improving run time by a factor
of about $30$).
The above example grammar for $t$ was produced 
by TreeRePair.
Note that the rank of a grammar is important, because it 
influences the run-time of algorithms that directly execute
on the grammar, such as executing tree automata or 
Core XPath~\cite{lohman06}.
Both BPLEX and TreeRePair take a user specified 
``maximal rank number $m$'', and produce grammars of
rank $\leq m$. 

\section{Structural Self-Index}

We call our XML self-index 
``TinyTree'' or simply ``TinyT''.
The first layer of storage in TinyT consists of a 
small memory representation of the grammar.
The second layer consists of additional 
mappings that support fast XPath evaluation.

\subsection{Base Index}
\label{sect:base}

The base index consists of a small memory representation of the
grammar. Start production right-hand sides are usually large
trees (they represent the incompressible part of the XML
structure tree) and are coded succinctly, using
two alternative ways. All other productions are transformed
into a normal form, so that each resulting production fits
into a single 64-bit machine word.
%The grammar compressors DAG, BPLEX, and
%TreeRePair, all produce grammars for which the initial
%start production is large (between 50k and 100k nodes for
%our documents), while
%all other productions are small (at most size $10$).
We experimented with two variants of representing the
start rhs:
\begin{enumerate}
\item[(bp)]
the succinct trees of Sadakane and Navarro~\cite{sadnav10}
\item[(ex)]
a naive custom representation.
\end{enumerate}
Both of these use $s\lceil\log\sigma\rceil$ many bits
to represent the tag sequence of the tree, where
$s$ is the number of nodes in the start rhs.
The first one uses the ``moderate size'' trees
of~\cite{sadnav10}, requiring additional 
$2s+O(s/polylog(s))$ bits of  space.
Our implementation of (bp) uses approximately $2.5$ bits per node.
The second one (ex) stores an explicit mapping 
called ``find-close'' which for 
every node records the number of nodes in its subtree.
This is sufficient for our XPath evaluators because they 
only
need pre-order access to the grammar, plus, the ability to
``skip'' a subtree. The find-close table allows to skip a 
subtree, by simply moving ahead in the tag-list by the number
of nodes specified in the table.
It requires $s\lceil\log s\rceil$ bits. Clearly, this
is rather wasteful compared to (bp), see the third column in 
Figure~\ref{fig:index_breakUp}), but can make a large
speed difference: e.g.,
our XPath count evaluator 
for the query Q06 $=$ /site/regions/*/item over XMark1G 
takes 3.5ms with (ex) and 4.7ms with (bp).
Observe also the difference in loading time of the
two variants shown in Figure~\ref{fig:index_times}.

We bring the remaining productions into
\emph{binary Chomsky Normal Form} (bCNF). 
A production is in bCNF if
it contains exactly two non-parameter nodes in its right-hand side.
The bCNF can be obtained following exactly the same procedure
as for ordinary CNF, see~\cite{lohmansch09}.
A grammar \emph{is in bCNF}, if every production except the
start production is in bCNF.
For instance, in our example cft grammar of before,
the $A$-production is not in bCNF.
We first change its right-hand side to 
$f(y_1, B)$ (which \emph{is} in bCNF) and add the
new production
$B\to a(c,c)$. The latter is not in bCNF and
therefore is changed to
$B\to C(c)$. The final grammar, called $\mathcal{G}_1$,
is:
\[
\begin{array}{lcl}
S&\to& A(A(a(b, c)))\\
A(y_1)&\to& f(y_1, B)\\
B&\to&C(c)\\
C(y_1)&\to& a(y_1, c)
\end{array}
\]
Note that the size of this grammar is $9$, thus has grown by one.
In general, the size of a grammar can grow by a factor $r$,
where $r$ is the rank of the original grammar.
The rank of the grammar can grow by $\mbox{max}(r,1)$, 
and the number of nonterminals can become at most
two times the size of the original grammar,
as implied by Proposition~3 of~\cite{lohmansch09}.
If we transform the DAG grammar for
$t$ of before into bCNF, then a grammar is obtained 
of rank one and of size $9$; consider
$t'=f(t,a(c,c))$, then the minimal DAG in bCNF is of
size $11$ (because two edges are added in the start rhs),
while our cft grammar has size $10$ (we simply add
another $A$-node in the start production).
In practice, we do not observe a large size increase;
the largest was $79$\%, see the last
column of Figure~\ref{fig:bCNF}.
\begin{figure}[ht!]
\centerline{
{\small
\begin{tabular}{|l|*{6}{@{\hspace{3pt}}c@{\hspace{3pt}}|}}
\hline
& size &\#rules & \#rules & depth & depth & size-diff\\
& start-rhs & (before) & bCNF & (before) & bCNF & (in \%)\\
\hline
XMark116M & 88299 & 10738 & 39631 & 36 & 361 & 21 \\
XMark1G & 64313 & 31684 & 284944 & 44 & 10381 & 79 \\
XMark11G & 105893 & 62604 & 408485 & 47 & 3915 & 70\\
Treebank83M\!\!\! & 470568 & 35352 & 37540 & 13 & 20 & 0.1\\
Sprot437M & 246970 & 20410 & 23484 & 128 & 165 & 0.3 \\
\hline
\end{tabular}
}
}
\caption{Impact of bCNF}
\label{fig:bCNF}
\end{figure}
Depth increase can be large as shown in the figure.
The rhs of each bCNF rule if of the form
$X(y_1,\dots,y_{i-1},Y(y_i,\dots,y_j),y_{j+1},\dots, y_r)$
and thus is characterized by the triple $(X,i,Y)$,
where $X$ and $Y$ are nonterminals or terminals, and $i$ is a number
between $1$ and the rank of $X$. 
We represent one bCNF rule by a single
64-bit machine word, using $28$ bits per nonterminal, $4$ bits
for the number $i$, and $4$ bits for the rank of the
nonterminal.
Our experiments show that setting the maximal rank of
BPLEX and TreeRePair to $8$ and $2$, respectively, 
gave best results for our XPath evaluators over the
corresponding indexes.
Thus, limiting our memory representation to grammars of rank $15$
($4$ bits) is justified.
We are now ready to calculate the \emph{space requirement of 
the grammar representation}:
%\[
$\#\text{CNF-rules} \cdot 8 \mbox{Bytes} +
\mbox{space}(\mbox{start-rhs})$.
%\]

As an example, for XMark116M we calculate, 
according to Figure~\ref{fig:bCNF},  $39631$ 
productions in bCNF, multiplied by $8$ bytes 
equals $309.6$KB.
For the tag sequence of the start rhs we
need $88299\cdot\lceil\log 39631 + 89\rceil=172.5$KB
(there are $89$ different labels for XMark).
For (bp) our implementation uses $27$KB, while (ex)
uses $183.2$KB.
Thus, the total sizes for (bp) and (ex) are,
$509$KB and $665$KB, respectively (the 
sum of the first three columns in Figure~\ref{fig:index_breakUp}, 
up to rounding).

\begin{figure}[ht!]
\centerline{
{\small
\begin{tabular}{|l|*{6}{@{\hspace{5pt}}c@{\hspace{5pt}}|}}
\hline
&CNF & STag & bp/ex & jump & prMap & SSkip\\
\hline
XMark116M & 310  & 173 & 27/183 & 431 & 305 & 345\\
XMark1G & 2226 & 149 & 20/126 & 3096 & 2222 & 251\\
XMark11G & 3191 & 246 & 32/220 & 4438 & 3403 & 414\\
%\hline
Treebank83MB & 293 & 919 & 144/1091 & 1178 & 260 & 1838\\
Sprot437M & 183 & 452 & 75/543 & 155 & 260 & 965\\
\hline
\end{tabular}
}
}
\caption{Sizes of TinyT's components (in KB)}
\label{fig:index_breakUp}
\end{figure}

\subsection{Auxiliary Indexes}
\label{sect:Aux}

There are two well-known principles of XPath optimization:
(1) jumping and (2) skipping. 
Here, \emph{jump} means to omit internal nodes of the document tree.
In our setting, the ``jumped'' nodes will be those represented by
a nonterminal of our grammar. We also say that the nonterminal
is ``jumped''. Note that after a jump we still need to continue
evaluating in the subtrees below the jumped pattern.
\emph{Skipping} means to omit a complete subtree. Thus, no 
evaluation is needed below skipped nodes.
We now introduce the jump table which allows to jump nonterminals; 
this table suffices for our XPath count evaluator.
To jump or skip 
during serialization and materialization we need further tables
(the ``pre and text mappings'').
Lastly, we mention the start-skip table which supports fast
skipping of subtrees.

\subsubsection*{Jump Table}
%\paragraph*{Jump Table}
As mentioned in the Introduction, it was observed 
in~\cite{arr+10,DBLP:journals/pvldb/ManethN10} that the two
operations taggedDesc and taggedFoll allow drastic speed-ups
for XPath evaluation. The SXSI system~\cite{DBLP:journals/pvldb/ManethN10}
keeps a large data structure (about 2.25 times larger than
the rest of their tree store) in order to give constant time access
to these operations. We try to add very little extra space to
our (so far tiny) index, and still be able to profit from
these functions. 
We build a ``jump table'' which keeps
for every nonterminal $X$ and every terminal symbol $b$
a bit indicating whether or not 
$X$ generates a $b$-labeled node.
When executing a taggedDesc-call (with label $b$), we try to derive
the first descendant node with label $b$; if a nonterminal during
this derivation does \emph{not} generate $b$'s (according to
our jump table), then we do not expand it, but ``jump'' it (by moving to
its first parameter position). The taggedFoll function
is realized similarly.
For the sequential interface (Point~2 in the Introduction)
plugged into SXSI~\cite{arr+10}
our experiments show that the speed-up through 
taggedDesc/taggedFoll 
is comparable to the speed-up obtained in SXSI.
%Roughly speaking, our fc/ns evaluator is about four times
%slower than SXSI, and similarly, our jump-evaluator is about
%four times slower than the jump-evaluator of SXSI.
This is surprising, because the space overhead 
for our jump table is small: $65\%$ of
extra space, compared to the $225\%$ in SXSI.

The jump table is not only useful to realize taggedDesc and
taggedFoll, but also allows 
speed-ups in all our XPath evaluators, see e.g.
%(which work over the
%grammar directly and do not use taggedDesc/taggedFoll).
%For instance, 
q1 in Figure~\ref{fig:simple}.
%goes from
%$22.4$ms to $3.0$ms (over XMark1G) when the jump table is used
%by the count evaluator.
The size of the jump table (in bits) is the number
of nonterminals multiplied by the number of different (terminal) labels.
For instance, XMark uses $89$ labels;
thus, the jump table for XMark116M is
39631 * 89 bits = 431KB.
For Treebank83M which has $257$ labels we obtain
37540 * 257 bits = 1177.7KB, see the fourth column
in Figure~\ref{fig:index_breakUp}.
In fact, our XPath count evaluator only loads the
base index plus the jump table, which implies the
total index sizes as shown in Figure~\ref{fig:TotIndexSizes} 
as sum of the 
first four columns in Figure~\ref{fig:index_breakUp} (taking ``ex'').

\subsubsection*{Pre and Text Mappings}
%\paragraph*{Pre and Text Mappings}
In order to be able to materialize pre-order node numbers,
or to access the text collection (needed for serialization),
we need to calculate, during evaluation, the numbers
of nodes/texts that have appeared until the current node
following a dflr traversal. However, if we ``jump'' a nonterminal
using our jump table, then we do not see its terminals.
Therefore we need another table which records for each
nonterminal the number of element nodes that it generates,
and similarly for the number of text nodes.
In fact, the situation is more complicated:
we actually need to store \emph{several numbers} per nonterminal,
as many as the rank of that nonterminal, plus one.
With respect to evaluation in dflr order, jumping a
nonterminal means to move
to its first parameter position and continue evaluation there.
Thus, we must know how many element
symbols are on \emph{the path from $t_X$'s root to the
first parameter}, where $t_X$ is the tree generated by $X$;
note that $t_X$ contains exactly one occurrence of each
parameter of $X$.
Similarly, once returned from $X$'s first parameter
position, we will want to jump to the second parameter.
We thus need to know the number of element nodes that are on the
path between $y_1$ and $y_2$ in the tree $t_X$.
The size of the corresponding table ``prMap'' is
$\sum_{X\in\text{\it NT}}(\text{\it rank\/}(X)+1)*\lceil\log k\rceil$,
where $k$ is the maximal number of element nodes on such paths, for any
nonterminal.
In fact, in our implementation we simply use a $4$-Byte integer
per value.
For instance, our grammar for 
XMark116M has 14057 nonterminals of rank zero,
14475 of rank one,
9311 of rank two, and
1786 of rank three. The size of the resulting prMap is
$(14057+14475*2+9311*3+1786*4=78084)*4\text{B}=305\text{KB}$;
this explains the column ``prMap'' 
in Figure~\ref{fig:index_breakUp}.
The corresponding table with numbers of text nodes
is called ``textMap'' table.

\subsubsection*{Start-Skip Table}
%\paragraph*{Start-Skip Table}
If, during materializing or  serializing
we want to skip a subtree, then we still need to traverse
that subtree of the grammar, in order to sum all numbers
of element nodes/texts, respectively (using the pr and text mappings).
To short-cut this
calculation, the \emph{start-skip table} is added. It stores
for every node of the start rhs, the total number of element nodes/texts
in its subtree. The size of this table is the number of
nodes in the start rhs multiplied by $\lceil\log n\rceil$,
where $n$ is the total number of element nodes/texts. 
In our implementation we simply use $4$ bytes per such number.
%For instance, the start rhs of the grammar for XMark116M 
%has $88229$ many nodes, thus 
%$88229 * 4\text{B} = 344\text{KB}$ are needed for
%the start-skip table, as shown in column ``SSkip'' in
%Figure~\ref{fig:index_breakUp}.
The corresponding table for the numbers of text nodes
is called ``textSSkip''.

\subsection{Index Generation}

The generation of the base index consists of the following
steps
\begin{enumerate}
\item[(1)] 
%(1) 
generate XML structure tree (MakeSTree),
\item[(2)] 
%(2) 
compress via TreeRePair,
\item[(3)] 
%(3) 
transform into bCNF, and
\item[(4)] 
%(4) 
build in-memory representation of TinyT components and save to
  file (BuildTinyT).
\end{enumerate}
Technically speaking, Steps (1) and (3) are not necessary
but can be incorporated into the TreeRePair compressor. 
In Step (1) we merely replace all text and
attribute values by placeholder nodes. This can be
incorporated into the parsing process of TreeRePair.
Similarly, TreeRePair can be changed so that it produces grammars
that are already in bCNF.
Since we also wanted to experiment with other compressors such
as DAG and BPLEX, we implemented small programs for (1) and (3). 
Our program for (1) is a naive java implementation using SAX
which is quite inefficient. 
Therefore the times for MakeSTree in Figure~\ref{fig:index_times}
should be ignored and the table should be read as:
indexing time is dominated by grammar compression time.
%takes time proportional to the grammar compression,
%plus a few seconds to build the actual index (bCNF and BuildTinyT). 
The times in Figure~\ref{fig:index_times} for step (4) are
for generating the base plus the jump index, i.e.,
the first four columns of Figure~\ref{fig:index_breakUp}.
The time for generating the two additional tables
prMap and SSkip (columns 5 and 6 in Figure~\ref{fig:index_breakUp}) 
is negligible, as it is is proportional to a ``chunk-wise'' traversal 
of the grammar (see Sections~\ref{sect:chunk}
and~\ref{sect:traversal}).

\begin{figure}[ht!]
\centerline{
{\small
\begin{tabular}{|l|*{4}{@{\hspace{2pt}}c@{\hspace{2pt}}|}}
\hline
& XMark116M & XMark1G & XMark11G & Treebank83M\\
\hline
MakeSTree & 0:26 & 3:45 & 38:38 & 0:30\\
TreeRePair & 0:33 & 8:58 & 52:10& 1:14\\
bCNF & 0:01 & 0:08 & 0:07 & 0:03\\
BuildTinyT & 0:01 & 0:04 & 0:06  & 0:02\\
\hline
Total  & 1:01 & 12:55 & 91:01 & 1:49\\
\hline
Peak memory & 182M & 728M & 2707M & 336M\\
\hline
Loading\hspace*{1.15mm} (bp) & 44ms & 45ms & 62ms & 209ms\\
\quad\qquad\quad (ex) & 2ms & 14ms & 18ms & 4ms\\
\hline
\end{tabular}
}
}
\caption{Times (min:sec) for index generation and loading}
\label{fig:index_times}
\end{figure}

\section{The Three Views of a Grammar}

An SLT grammar can be seen as a factorization of a tree
into its (repeated) tree patterns. Each tree pattern
is a connected subgraph of the original tree and is represented
by a nonterminal.
In our algorithms we found a hierarchy of three different views
of the grammar:
\begin{enumerate}
\item[(1)] node-wise (the slowest),
\item[(2)] rule-wise (the fastest), and
\item[(3)] chunk-wise. 
\end{enumerate}
The \emph{node-wise view} is a proxy to the original tree and
allows to execute arbitrary algorithms (using the first-child,
next-sibling, and parent functions).
This is the most
``detailed'' view, but causes a slow-down (comparable to the
space improvement of the grammar, when compared to succinct trees).
The \emph{rule-wise view} is the most abstract and high-level view; 
it means to move through the grammar in one pass, rule by rule.
Specialized algorithms such as executing finite-state
automata can operate in this view. For strings this
idea is well studied~\cite{ryt04_short}. 
%and
%belongs to the well-known topic of computing over compressed
%structures, see e.g.,~\cite{amibenfar96}. 
We show in Section~\ref{sect:rule-wise} that
the ``selecting tree automata'' of~\cite{DBLP:journals/pvldb/ManethN10} can
be executed in the rule-wise view in order to count selected nodes.
This is applied to XPath in Section~\ref{sect:XPath} by compiling
queries into selecting automata.
The \emph{chunk-wise view} is slightly more detailed than the rule-wise
view. It means that the grammar is traversed (once) in a strict 
dflr order. This allows to keep track of pre-order numbers and
text numbers, by keeping a global counts of element nodes/texts.
Through the prMap and SSkip tables we can apply jumping in
this view which allows to build fast XPath evaluators
for serialization and materialization.
Processing in the chunk-wise view is slightly slower than rule-wise
(proportional to the rank of the grammar), because 
the rule of a nonterminal of rank $k$ is now processed
$k+1$ times (instead of only once in rule-wise).
%In fact, in our experiments we observe a slow down of
%factor $3-4$ for all grammars (independent of their rank)
%of documents up to 1GB (including Treebank83M and Sprot437M), 
%and of factor $5$ for XMark11G.

\subsection{Node-Wise View}
\label{sect:node-wise}

The node-wise interface 
allows to execute arbitrary algorithms over the original tree
(see, e.g.,~\cite{buslohman08}).
%(mentioned already in~\cite{manbus04_short}).
Inside the interface, a node is represented by a sequence
of pairs which shows the productions that were applied
to obtain the node. 
The length of such sequences is at most the depth of the grammar,
which be as large as $10000$ (see Figure~\ref{fig:bCNF}). 
Thus, even if one pair fits 
into a single bit (which \emph{can} be done) then this is 
large compared to the 32 or 64 bits for a pre-order node ID.
We observe a slow-down of the original algorithm of approximately the
same factor as the compression.
Recursive tree algorithms need a lot of memory due to the size
of these sequences. For iterative algorithms we obtain very
good time/space trade-offs, see Figure~\ref{fig:travIt}.

The node-wise interface provides the functions 
find-root,
first-child,
next-sibling, and parent (plus checking the current label of
course). A node of the original tree is represented as a sequence
of pairs
$\eta = (\text{Start}, p_0)(A_1, p_1) \cdots (A_j, p_j)$,
where $p_0$ is a node of the start rhs,
$A_1,\dots, A_j$ are nonterminals,
and $p_1,\dots, p_j$ are nodes such that 
rhs(S) at node $p_0$ is labeled $A_1$, and for
every $1\leq i<j$,
the rhs for $A_i$ at node $p_i$ is labeled $A_{i+1}$. Moreover, it must
hold that the rhs of $A_j$ at node $p_j$ is labeled by a terminal
symbol, say $b$. The node ID $\eta$ is \emph{labeled by} $b$,
denoted $\text{lab}(\eta)=b$.
The first child (fc), next sibling, and parent functions are 
realized as in Section~6.2 of~\cite{buslohman08}. For instance,
$\text{fc}(\eta)$ is the following node ID:
we first move to the first child of $p_j$ in the rhs of $A_j$,
if it exists.
There are three possibilities: (1) fc($p_j$) is labeled 
by a terminal symbol. In this case we are finished and
return $\eta[(A_j,p_j)\leftarrow (A_j,p_j.1)]$, i.e.,
$\eta$ with the last pair replaced by $(A_j,p_j.1)$.
(2) fc($p_j$) is labeled by a nonterminal $A_{j+1}$.
Let $\eta'=\eta[(A_j,p_j)\leftarrow (A_j,p_j.1)](A_{j+1},\varepsilon)$.
If the rhs of $A_{j+1}$ has a terminal at its root node, then
the process is finished and return $\eta'$.
Otherwise, more nonterminals $(A_{j+1},\varepsilon)\dots
(A_{j+k},\varepsilon)$ are added until
$A_{j+k}$ has a terminal root node (and all
$A_{j+1},\dots,A_{j+k-1}$ do not). (3) fc($p_j$) is labeled by
a parameter $y_i$. We remove the last pair from $\eta$
and consider the $i$-th child of the node $p_{j-1}$ in
the rhs of $A_{j-1}$. If it is a terminal, then we are
finished. If it is a nonterminal, then we expand as in Step~2.
If it is again a parameter, then the pair is removed again,
until a non-parameter last pair is found.
This terminates with the desired node ID of the first-child node.

As an example, the node ID
$\eta_0=(S, \varepsilon)(A, \varepsilon)$ represents the $f$-labeled
root-node of the tree represented by our example grammar
$\mathcal{G}_1$. To compute 
$\text{fc}(\eta_0)$ we move to the first child of $f$ in
$A$'s rhs. This is the parameter $y_1$. Thus, we pop $\eta_0$
and move to the second $A$ of the start rhs,
$(S, 1)$. We expand the $A$ in one step and obtain
the result $(S, 1)(A, \varepsilon)$. 

\subsection{Rule-Wise View}
\label{sect:rule-wise}

The rule-wise view means that the grammar is traversed only once, 
rule by rule, and in each step only little computation takes place which
is ``compatible'' with the grammar. 
A classical example of this kind of ``computing over compressed
structures'' is the execution of a finite-state automaton over a
grammar compressed string, i.e., over a straight-line context-free grammar
(see, e.g., Theorem~9 of~\cite{ryt04_short}).
The idea is to memoize the ``state-behaviour'' of each nonterminal.
For tree automata over SLT grammars, 
the problem was studied in ~\cite{lohman06}
from a complexity theory point of view.
We use \emph{selecting tree automata} as
in~\cite{DBLP:journals/pvldb/ManethN10} and build a 
``count evaluator'' which executes in one pass over the grammar. 
It counts the number of result nodes of the given XPath query. 

The new aspect is to combine this evaluator with 
the jump table.
Intuitively, if in a certain state only a
given label $b$ is \emph{relevant} (meaning that only for that
label the automaton changes state or selects the node), then
we can jump over nonterminals that do not produce this label $b$
(determined by the jump table).
For instance, consider the query //$f$//$b$ which selects all b-descendants
of f-nodes. It should be intuitively clear that this query
can be answered by considering only the $f$ and $b$-nodes
of the document (and their relationship).
This means that during top-down evaluation we may 
jump nonterminals which do not produce $f$ or $b$ nodes.
We now introduce, by example, \emph{selecting tree automata}
(\emph{ST automata}), and
discuss how they can be executed for result-counting over a grammar. 
We then show how jumping can be incorporated into this process.
Here is an example of an ST automaton:
\[
\begin{array}{lcl}
q_0, f&\to& q_1,q_0\\
q_0,L-\{f\}&\to& q_0,q_0\\
q_1,b&\Rightarrow&q_1,q_1\\
q_1,L-\{b\}&\to&q_1,q_1
\end{array}
\]
The first rule says that if in state $q_0 $ the automaton
encounter an $f$-labeled node, then it moves to state $q_1$ at
the first child, and to state $q_0$ at the second child.
The second rules says that, in state $q_0$ and 
for all labels (denoted by $L$) except $f$, it stays
in state $q_0$ at both children nodes. In state $q_1$
the current node is selected if it is labeled $b$ (denoted by
the double arrow '$\Rightarrow$' in the third rule). The automaton 
realizes the XPath query //$f$//$b$ over our binary tree
representation of XML trees.
We now want to execute this automaton over the grammar
$\mathcal{G}_1$ of Section~\ref{sect:base}, in ``counting mode'',
i.e., producing a count of the number of result nodes.
It starts in state $q_0$ processing the start rhs of the
grammar. Its root node is labeled $A$, so the automaton moves to
the $A$-production (still in state $q_0$).
The first automaton rule applies at the $f$-labeled node, meaning to
process the first child ($y_1$) in state $q_1$ and the second child  $B$
in state $q_0$.
The latter means to process $C$ in state $q_0$ which gives
state $q_0$ at $y_1$.
We are now finished with processing the nonterminal $A$ in state $q_0$.
In summary: no result node was encountered, and the state
has moved from $q_0$ to state $q_1$ at the first parameter $y_1$.
This ``behaviour'' of $q_0$ on $A$ is hashed as
$(0,q_1)$.
Of course, during this computation, the corresponding
behaviors for $C$ and $B$ are hashed too, i.e.,
for $q_0$ on $C$ the value $(0,q_0)$ and for 
$q_0$ on $B$ the value $(0)$.
The automaton continues in state $q_1$ at the second $A$-node of
the start rule. Unfortunately,
no hash for $q_1$ on $A$ exists yet, so the automaton needs to be run. 
Again no result node is encountered
and it stays in state $q_1$ at $y_1$.
Thus, $(0,q_1)$ is hashed for $q_1$ on $A$. 
Finally, it processes the $a$-node of the start
production, in state $q_1$. It gives $q_1$ at the $b$-node.
This node is selected according to the third rule
of the automaton and therefore our global result
count is increased, to its final value of one.
Observe that if there was a third $A$-node
in the start rhs, such as for the slightly larger tree $t'$
mentioned before, then hashing is already useful 
because there will be a hash-hit for the third $A$.
It should be clear that, in the same way, 
any ST automaton can be processed 
in one pass through the grammar (see also~\cite{fisman07,lohman06}).
Note that we only evaluate ST automata that are
\emph{deterministic}; it means that for every state $q$ and
every label $a$ there is at most one transition with 
left-hand side ``$q$, $a$''.
%Further, since running over the start rhs
%is not done by a recursive procedure, we maintain by hand a stack
%which stores the states in which further 
%children (on the path to the current node) must still need to be processed
%(similar to the ``PC stack'' of example in Section~\ref{sect:chunk}).
%We now investigate how to increase the efficiency of
%this procedure, through the use of our jump table.

\subsubsection*{Adding Jumping}
%\paragraph*{Adding Jumping}
Consider the example automaton of before.
%, 
%over the same grammar (displayed in Section~\ref{sect:base}).
It should be clear that in state $q_0$ the
automaton only cares about $f$-labeled nodes, i.e., it can
omit all other-labeled nodes and safely proceed to the first
$f$-labeled descendant node (if such a node exists).
In the terminology 
of~\cite{DBLP:journals/pvldb/ManethN10}, the omitable nodes are
``not relevant''. Here we say that a node is \emph{relevant} 
if the 
automaton either selects the node, or 
changes state, i.e., applies a transition
with rhs $(q',q'')$, where $q'\not=q$ or $q''\not=q$.
Note that 
in~\cite{DBLP:journals/pvldb/ManethN10} relevance is defined
based on minimal automata; we have dropped this restriction and
define it for arbitrary (but deterministic) ST automata.
We further say that for state $q$,
$u$ is a \emph{relevant label} if the automaton's
transition for $q$ and $u$ is selecting, or 
changes state, i.e., has rhs $(q',q'')$ with
$q'\not=q$ or $q''\not=q$.
Obviously, during the run of the automaton, the relevant labels
allow to determine the next relevant node.

We use the jump table in order to omit (``jump'') nonterminals
which do not contain relevant nodes for the current state $q$:
if the jump table indicates that a nonterminal does
\emph{not} produce nodes labeled $U=u_1,\dots,u_k$, and
the relevant labels of the current state are in $U$,
then the nonterminal may be jumped. By our definition of
relevance this implies that all parameters of of jumped nonterminal
will all be processed in state $q$.
Back to the example:
Since $f$ is a relevant label for $q_0$, we cannot jump the 
first $A$-node of the start rhs. Hence, the automaton proceeds as before and
eventually the entry $(0,q_1)$ is hashed for $q_0$ and $A$.
The automaton proceeds 
in state $q_1$ at the second $A$-node of the start rhs.
The only relevant label for $q_1$ is $b$.
The jump table tells us that $A$ \emph{does not generate} $b$'s. 
Thus, we jump this $A$ and continue evaluating at its child node.
This saves a lot of computation (roughly half of before).
But, in which state is the automaton supposed to continue?
It must be state $q_1$ because, by 
definition of relevance, the state never changes
on all non-relevant nodes. Thus, 
parameter $y_1$ must be reached in state $q_1$.
We proceed, to the $b$-node of the start rhs
and compute the correct final count of $1$.

As another example, 
imagine the start rhs was $A(A(b))$ and we execute 
a query that selects all $b$-children of the root node.
In XPath /$b$
(let us ignore that in XML the root node has only one child).
An automaton for this query is:
\[
\begin{array}{lcl}
q_0,b&\Rightarrow&q_1,q_0\\
q_0,L-b&\to &q_1,q_0\\
q_1,L&\to& q_1,q_1.
\end{array}
\]
In state $q_0$, all labels are relevant, because there
is a state change in all transitions for $q_0$.
We therefore process as before,
eventually hash the entry $(0,q_1)$ for $q_0$ and $A$,
and determine that $y_1$ of $A$ need to be processed in
state $q_1$.
For this state, no label is relevant. Thus, the second $A$
may be jumped. We arrive at the b-node of the (new) start
production of above, and terminate (with count zero).

\subsubsection*{XPath Specific Finer Relevance}
%\paragraph*{XPath Specific Finer Relevance}
For XPath, we found it beneficial to use 
a slightly finer definition of relevance. It allows to jump more
nonterminals for automata that realize XPath.
First, define that 
a state $q$ is \emph{universal} if it has the transition $q,L\to(q,q)$,
i.e., never changes for any label.
For all our automata there is at most
one (fixed) universal state which is denoted
by $q_U$. For instance, in the above automaton for /$b$, $q_U=q_1$.
\iffalse
We define:
a node is \emph{f-relevant} 
if the automaton either selects the node, or 
applies a transition with rhs $(q',q'')$,
where 
($q'=q''=q_U$) or
($q'\not=q$ and $q'\not= q_U$) or $q''\not=q$.
For state $q$ the label $u$ is \emph{f-relevant} if
the $(q,u)$-transition is selecting,
or its rhs is of the form $(q',q'')$ with
($q'=q''=q_U$) or
($q'\not=q$ and $q'\not= q_U$) or $q''\not=q$.
\fi
We define:
a node is not \emph{f-relevant} 
if the automaton does not selects the node, and
applies a transition with rhs $(q,q)$,
$(q_U,q)$, or $(q_U,q_U)$.
For state $q$ the label $u$ is not \emph{f-relevant} if
the $(q,u)$-transition is not selecting,
and its rhs is of the form $(q,q)$,
$(q_U,q)$, or $(q_U,q_U)$.
%tom's def of non-f-relevance: In a state $s$ a label $l$ is not \emph{f-relevant} if the automaton $\mathcal{A}=(\mathcal{L},\mathcal{Q},\iota,\delta,\mathcal{M})$ does not mark it, that is $(s,l)\notin\mathcal{M}$, and if $\mathcal{A}$ applies a transition of the form $\delta(s,l)=(s,s)$, $\delta(s,l)=(\mathcal{U},s)$, or $\delta(s,l)=(\mathcal{U},\mathcal{U})$.
Let us consider the last example of above again,
the automaton for /$b$ over our example grammar.
This time, only $b$ is a relevant label for $q_0$, because
$q_U=q_1$. 
The rule for jumping non-$f$-relevant nodes, in a given state $q$, is:
the $q$-transitions for all non-$f$-relevant labels must
all have the same rhs, which, itself is one of
$(q,q)$, $(q_U,q)$, or $(q_U,q_U)$.
Thus, we may jump the first $A$-node
of the start rhs.
Now it is more difficult to determine in which state 
to proceed at $y_1$ of $A$:
the root node of $A$'s pattern tree, and its descendants
of the form 2.2.2$\cdots$.2 (in Dewey notation) are
processed in state $q_0$, while all other nodes are
processed in state $q_1$. 
Since $A$'s pattern tree is $f(y_1,a(c,c))$, this
means that $q_1$ is the correct state for $y_1$.
However, if $A$'s pattern tree was different,
for instance $f(a(c,c),y_1)$, then
we would need to assign the state $q_0$ to $y_1$.
This shows that in order to correctly
jump a nonterminal $X$ which contains no $f$-relevant nodes,
we need to statically know 
whether or not $X$'s last parameter $y_j$ occurs at a $2.2.\dots.2$-node
in $X$'s pattern tree $t_X$. 
This information is determined at indexing time and is 
stored with the grammar as part of our index. 
Since its size is negligible
(one bit per nonterminal), we do not
explicitly mention it in our size calculations.
%;
%for instance, we need $50$KB extra space to store this
%information for the index of XMark11G.

\subsubsection*{Adding Skipping}
%\paragraph*{Adding Skipping}
When an automaton is in its universal state $q_U$,
we may skip the entire subtree because it 
contains no relevant nodes.
For the count evaluator this is done by omitting
all recursive calls to state $q_U$. This holds for
terminal nodes, as well as for the hashed behavior
of nonterminal nodes. 
For the materialize and serialize evaluators, 
it is necessary to know the number of element nodes/text nodes 
of the skipped subtree to correctly continue evaluating.
During recursion these numbers are determined by
the prMap/textMap tables. If we are in the start rhs,
then we use the SSkip/textSSkip tables.

\subsection{Chunk-Wise View}
\label{sect:chunk}

%Instead of counting the number of selected nodes 
We now wish
to \emph{serialize} XML result subtrees of the nodes selected by
an automaton. Additional to the grammar, we need access to
the text values of the XML document. We assume a function 
getText($i$) which returns the $i$-th text or attribute value
of the document (starting from zero). 
For instance, getText($6$) returns the $7$-th text value, i.e, the
string \texttt{serialization} for this example document

\begin{verbatim}
<g>This<f><f><a><b>is</b></a><c>a test</c></
f><a><c>document</c><c>for the purpose</c></
a></f><a><c>of explaining</c><c>serializatio
n</c></a></g>
\end{verbatim}

A faithful grammar representation of the XML structure tree of
this document is:
\[
\begin{array}{lcl}
S&\to& g(\_T, A(A(a(b(\_T), c(\_T)))))\\
A(y_1)&\to& f(y_1, B)\\
B&\to&C(c(\_T))\\
C(y_1)&\to& a(y_1, c(\_T))
\end{array}
\]
For simplicity we do not transform this grammar into bCNF.
We would like to serialize (using the jump table)
the nodes selected by this automaton:
\[
\begin{array}{lcl}
q_0,c&\Rightarrow&q_0,q_0\\
q_0,L-c&\to &q_0,q_0
\end{array}
\]
The automaton begins in state $q_0$ at the root of
the start rhs. The recursive algorithm over grammar rules
is shown in Figure~\ref{fig:print}; exactly the same algorithm is used
over the start rhs (but iteratively, using stacks).
During a dflr traversal the global counter num\_T stores
the number of \_T nodes seen so far.
Thus, at $g$'s first child $\text{num\_T}$ is set to $1$.
We now process, still in state $q_0$, 
$A$'s \emph{first chunk} (that is: the sequence of tags
from $t_A$'s root node to its first parameter node).
This is done by first calling the rule-wise evaluator
of Section~\ref{sect:rule-wise}, in order to compute and
hash the parameter states for $A$ and the information
whether a parameter is inside a result subtree (see the $a_i$'s
in the algorithm of Figure~\ref{fig:print}. 
This will add the
hashes $(q,q_0)$ for $q_0$ on $C$, and
$(2)$ for $B$, and $(2,q_0)$ for $A$.
The first chunk only contains $<$f$>$ and therefore
the empty tag sequence $(0,0)$ is hashed for 
$A$'s first chunk in $q_0$, i.e., for the 
triple $(q_0,A,1)$. Further, $(q_0,A,2)$ is pushed onto
our ``pending computation'' (PC) stack.
The next step in dflr is the first chunk of the second $A$-node.
Both rule-wise and chunk-wise behaviors are hashed already,
so nothing needs to be computed and again
$(q_0,A,2)$ is pushed onto the PC stack.
The dflr traversal continues at the subtree
$a(b(\_T),c(_\_T))$. 
The $a$ and $b$ nodes do not cause state changes
or node selection.
At the first \_T-node, num\_T is set to $2$.
At the $c$-node a selecting transition fires.
Thus, we now start appending tags to the
``intermediate result tag'' (IRT) sequence, first
the tag $<$c$>$. 
We also append the
pair of start position and current num\_T value to
the ``final result list''. 
Moreover, $<$/c$>$ is pushed onto the PC stack. 
Evaluation continues at the \_T-child (thus num\_T is increased to $3$). 
We append $<$\_T/$>$ to the IRT sequence. Since \_T is a leaf,
the PC stack is popped and therefore add
$<$/c$>$ to the IRT sequence. 
This finishes the result
subtree. 
At the next step we
return to the $a$-node of the start rhs. We return
to the second $A$-node and pop the PC stack. This gives
$(q_0,A,2)$. No $($begin, end$)$-pair is hashed for this triple,
so
$A$'s second chunk is processed in state $q_0$.
Recursion continues to $B$ and $C$ and finally move to the parameter 
tree $c(\_T)$
of $C$ in $B$'s rhs. This causes to append
$<$c$>$$<$\_T$>$$<$/c$>$ to the IRT sequence,
to append $(4,2)$ to the final result list,
and to increase num\_T (to $4$).
We proceed at the second chunk of $C$,
ignore $<$/a$>$, and 
append $<$c$>$$<$\_T$>$$<$/c$>$ and $(7,3)$ to the
IRT sequence and final result list, respectively, 
and increment num\_T (to $4$).
The pair $(7,9)$ is now hashed for the triple $(q_0,C,2)$.
The grammar recursion continues at $B$ and $A$, so
$(4,9)$ is hashed for $(q_0,B,1)$ and
$(4,9)$ for $(q_0,A,2)$. The dflr run continues at the
first $A$ of the start rhs and pop the PC stack.
This gives $(q_0,A,2)$.
We now have our first hash-hit and happily retrieve
the (begin,end)-pair $(4,9)$. This is interpreted as
a ``copy instruction'': append to the IRT sequence 
(currently at position $10$) its
own content from position $4$ to $9$.
During this copying we observe that $4$ and $7$ are final
result begin-positions, and that their corresponding
num\_T-values are $5$ and $6$, respectively
(by incrementing num\_T during copying).
%We thus add $(10,5)$ and $(13,6)$ to the final result list.
%Evaluation is now finished. 
The content of the final result list is
$(1,2)(4,3)(7,4)(10,5)(13,6)$.
The IRT sequence contains five copies of
$<$c$>$$<$\_T$>$$<$/c$>$.
In a final step we print correct XML document fragments
for each result. This is done by copying from the IRT sequence
while inserting for each $<$\_T/$>$ the correct text value.

\begin{figure}[ht!]
\begin{framed}
{
\small
\begin{tabular}{l}
\textbf{function} \emph{recPrint}(nt $N$, state $s$, chunkNo $p$, bool 
$u$) $\{$ \\
\textbf{let} $S$ = ($X_1$, $s_1$, $p_1$, $u_1$)($X_2$, $s_2$, $p_2$, $u_2$) 
$\dots$
($X_n$, $s_n$, $p_n$, $u_n$) \\
\quad\quad be the  of T/NT-chunks in rhs($N$) between ``$y_p$ and
$y_{p+1}$''; \\
int currLength = IRT\_length; \\
 \textbf{for} $i$ = $1$ \textbf{to} $n$ \textbf{do} \\
\quad \textbf{if} ($X_i$ = nonterminal) \textbf{then} \\
\quad\quad \textbf{if} (hash$(X_i$, $s_i$, $p_i$, $u_i)$ = ($z_1$, $z_2$))
\textbf{then}\\
\quad\quad\quad \textbf{for} $j$ = $1$ \textbf{to} $z_2$ \textbf{do} \\
\quad\quad\quad\quad append(IRT, IRT[$z_1$ + $j$]);\\
\quad\quad\quad\quad \textbf{if} (IRT[$z_1$ + $j$] is a result)
\textbf{then} \\
\quad\quad\quad\quad\quad append(FRL, (IRT\_length, num\_T));\\
\quad\quad \textbf{else} \emph{recPrint}($X_i$, $s_i$, $p_i$, $u_i$); \\
\quad \textbf{else} \\
\quad\quad \textbf{if} ($p_i$ = $0$) \textbf{then} \\
\quad\quad\quad \textbf{if} (($s_i$, tag($X_i$)) is selecting \textbf{or} $u_i$ = $1$)
\textbf{then}\\
\quad\quad\quad\quad append(IRT, ``<$\text{\it tag\/}(X_i)$>'');\\
\quad\quad\quad \textbf{if} (($s_i$, tag($X_i$)) is selecting) \textbf{then}\\
\quad\quad\quad\quad append(FRL, (IRT\_length, num\_T));\\
\quad\quad\quad \textbf{if} ($tag(X_i)$ = $\_T$) \textbf{then}
num\_T++;\\
\quad\quad \textbf{if} ($p_i$ = $1$ \textbf{and} (($s_i$, tag($X_i$)) is selecting
\textbf{or} $u_i$ = $1$) \textbf{then} \\
\quad\quad\quad append(IRT, ``$<$/$\text{\it tag\/}(X_i)$$>$'');\\
 hash($N$, $s$, $p$, $u$) = (currLength, IRT\_length - currLength);\quad$\}$\\
\end{tabular}
}
\caption{Grammar-recursive case of the print function}
\label{fig:print}
\end{framed}
\end{figure}
To see how jumping works, consider the query //$b$ over this grammar.
Now $A$'s first chunk can be jumped. 
%At this moment we need to consult the prMap (in fact, the text
%version of it, let us call it textMap) to correctly increment
%the num\_T-count (in this case, by zero; so it remains $1$). 
During the rule-wise traversal, jumping takes place as discussed
in Section~\ref{sect:rule-wise}. Next, the first chunk of the second
$A$-nods in the start rhs is jumped. The first hit for //b
is obtained at the $b$-node of the start rhs. The dflr traversal
jumps the second chunks of both $A$-nodes of the start rhs, and 
is finished. The final result list is $(1,1)$ and
the IRT sequence is $<$b$>$$<$\_T$>$$<$/b$>$.
Thus, we print \verb!<b>is</b>!.

Comments to Figure~\ref{fig:print}:
In Line~2 we calculate \emph{rule-wise} the parameter states
$s_1, \dots, s_n$ and the Booleans $u_1, \dots, u_n$ 
which determine if a parameter is inside of a result
subtree. Line~3:
if $p=0$ then $y_p$ refers to the root node and if 
$p=\text{rank}(N)$ then $y_{p+1}$ also refers to the root node.
The $X_i$ and $p_i$ are determined by the shape of $\text{rhs}(N)$.

\section{XPath Evaluation}
\label{sect:XPath}

We built rudimentary XPath evaluators 
that compile a given XPath query into an ST automaton.
Our current evaluator only works for the /, //, and following-sibling
axes and does not support filters. 
The count evaluator is based on the rule-wise view
of Section~\ref{sect:rule-wise} while the materialize and
serialize evaluators are based on the chunk-wise view
of Section~\ref{sect:chunk}.
For the small XPath fragment considered here, the translation
into ST automata is straightforward and similar to the one
of~\cite{arr+10,DBLP:journals/pvldb/ManethN10}
(essentially, the automaton is \emph{isomorphic} to the query).
First, an automaton is built which uses nondeterminism
for the //-axis. For instance, we first obtain an 
automaton similar to the one shown in the
beginning of Section~\ref{sect:rule-wise}, but
with $L - \{f\}$ replaced by $L$, and
$L - \{b\}$ replaced by $L$. 
Different from~\cite{arr+10,DBLP:journals/pvldb/ManethN10} which
work on-the-fly, we fully determinize the automaton before evaluation. 
For the example, this gives precisely the automaton as shown. 
We can prove that determinization of our ST automata does not
cause an exponential blow up. This is due to the simple form
of our queries. Moreover, the determinization procedure always
produces minimal automata.
In fact, it can be shown that for a given XPath query 
with $m$-axes (over /, //, and following-sibling) the
resulting deterministic ST automaton has at most
$2m$ states.
Note that in terms of the transitions along a first-child path,
our deterministic ST automata behave exactly in the same
way as ``KMP-automata'' (see, e.g., Chapter~32 
of~\cite{DBLP:books/mg/CormenLRS01}), i.e., matching along
a path works very much in the same way as the well-known
KMP-algorithm.
What is the time complexity for counting, i.e., 
of executing a deterministic ST automaton over an SLT grammar?
As mentioned already in~\cite{lohman06}, even for general context-free
tree grammars, a deterministic top-down tree automaton can be executed
in polynomial time. 
We make this more precise: for each nonterminal (of rank $k$)
of the grammar and state of the automaton, we need to compute only 
at most one $k$-tuple of parameter states. Hence, the following holds.

%\smallskip

%{\bf Lemma.}\quad
\begin{lemma}
Let $G$ be an SLT grammar in which every production is in bCNF
and let $M$ be an ST automaton.
Let $n$ be the number of nonterminals of $G$, $k$ the rank of $G$,
and $m$ be the number of states of $M$.
The automaton $M$ can be executed rule-wise (e.g., for counting)
over the grammar $G$ in time $O(mnk)$.
\end{lemma}

%\smallskip

Note that an alternative way is to first reduce the number
of parameters of the grammar to one, using the result 
of~\cite{lohmansch09}. For a binary ranked alphabet (as we
are using here for XML), the size of the resulting grammar
is $O(2 |G|)$, where $|G|$ denotes the size of $G$. 
We then apply the above theorem in time $O(mn')$, where
$n'$ is the number of nonterminals of the new grammar.
It remains to be seen in practice which of the two approaches 
give better running times.

\section{Experiments}
\label{sect:experiments}

All experiments are done on a machine featuring 
an Intel Core2 Xeon processor at 3.6GHz, 3.8GB of RAM, and
an S-ATA hard drive. The OS is a 64-bit version of Ubuntu Linux.
The kernel version is 2.6.32 and the file system is 
ext3 with default settings. All tests are run with only the
essential services of the OS running.
The standard compiler and libraries available on this
distribution are used
(namely g++ 4.4.1 and libxml2 2.7.5 for document parsing).
%\paragraph*{Protocol}
Each query is run three times and of the three running times
select the fastest one. We only count query execution time, i.e., 
do not take into account query translation times etc.
For experiments that involve serialization the programs
are directed to write to /dev/null. 
\paragraph*{MonetDB:}
We use Server version 4.38.5, release Jun2010-SP2.
This contains the MonetDB/XQuery module v0.38.5.
%For count queries, we found that MonetDB is faster
%over the original XMark documents, than over 
%documents with text and attributes removed.
%Inside {\small \texttt{mclient -l xquery -t}} we run
%{\small \texttt{count(fc:doc("XMark..")/query)} }
We compare pure query execution time, so
report the ``Query'' time reported.
% (and
%ignore all other times).
%For serialization, we redirect the output
%to /dev/null to get disk-independent
%times (the same is done with all other programs).
%We use the ``Timer'' value reported by 
%MonetDB for serialization.
%In comparison to the times reported in~\cite{arr+10} which
%were obtained with an older version of MonetDB, we observe
%that (except for Q01 and Q02) all queries run faster with
%the new version, expecially over larger documents.
%For some queries such as the
%``crash test'' queries Q13-Q16 the difference is
%significant ($5$--$10$-times faster).
\paragraph*{Qizx}
%\textbf{Qizx:}
Version 4.0 (June 10th, 2010) of the free engine is used.
The ``-v -r 2''-switches are used. For count queries we
use the ``evaluation time:''-number reported by Qizx.
For serialization the sum of
the ``evaluation time:'' and ``display time:''-numbers are used.
%In comparison to the times reported in~\cite{arr+10} which
%were obtained with an older version of Qizx, we observe
%that all queries run faster with the new version.
%For some queries the difference
%is significant, e.g., Q13 runs $17$-times faster with the
%new version. Only three queries became slower:
%Q05, Q07, and Q08, by up to $12$-times (Q07, serialization
%over XMark1G); this seems to be due to a change in 
%implementation for the //-axis of XPath.
For a few count queries Qizx executed faster over the
XML structure tree than over the original XML document.
This is indicated by a footnote in Figure~ref{fig:xmarkrun}.
\paragraph*{SXSI}
%\textbf{SXSI:}
The version used for~\cite{arr+10} was supplied to us by
the authors.

\subsection{Traversal Access}
\label{sect:traversal}

To investigate the speed of our node-wise view, 
we consider fixed traversals: depth-first left-to-right (dflr) and 
dfrl traversals, both recursively and iteratively.
Dflr traversals are common access pattern for
XPath evaluation. 
%The recursive traversal only
%needs find-root, fc, and ns, while the iterative one also needs
%the parent function.

\begin{figure}[ht!]
\vspace*{-1mm}
\centerline{
\includegraphics[width=8cm]{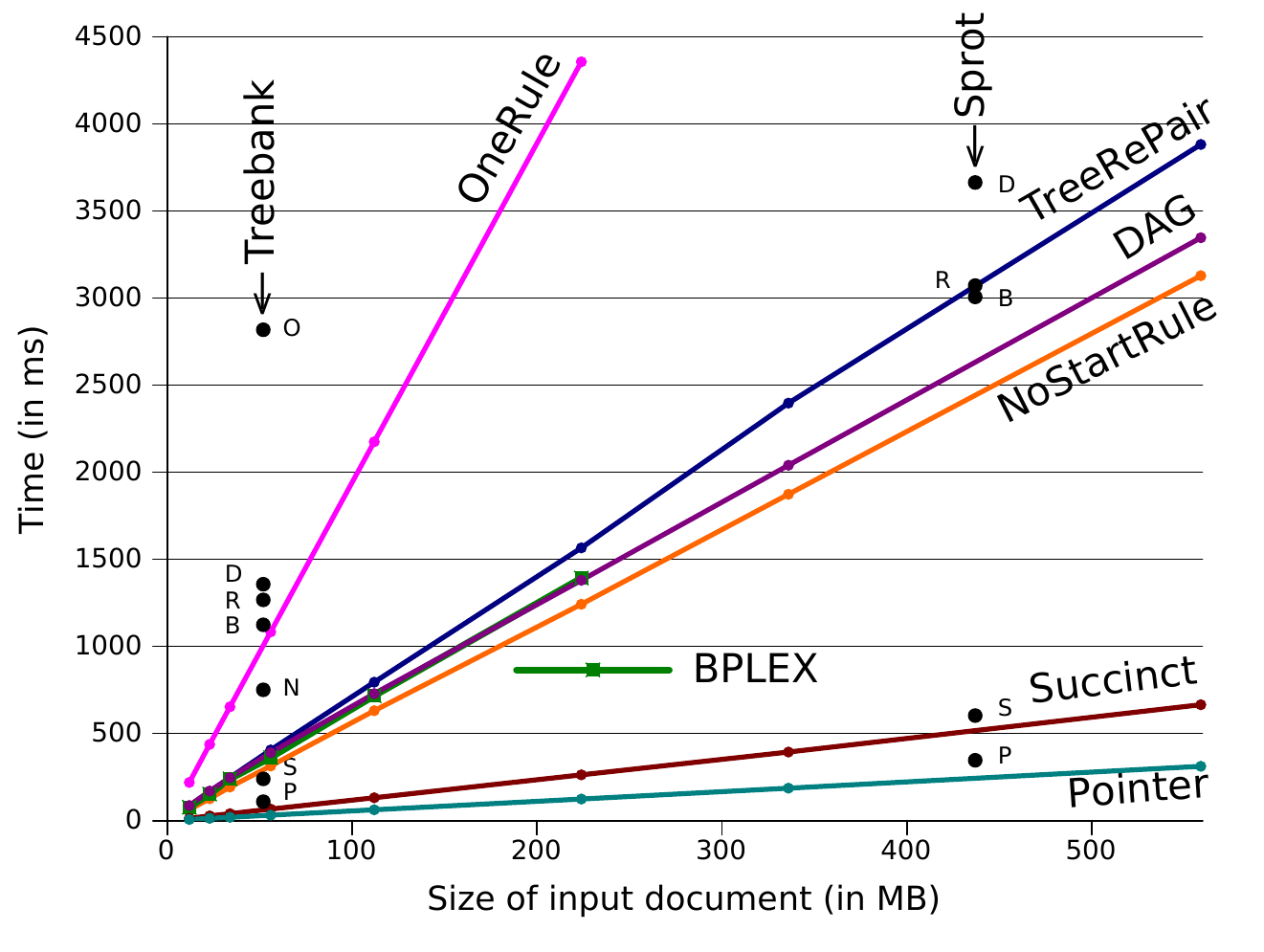}
}
\vspace*{-4mm}
\caption{Recursive tree traversals over XMark}
\label{fig:travRec}
\end{figure}

\begin{figure}[ht!]
\vspace*{-1mm}
\centerline{
\includegraphics[width=8cm]{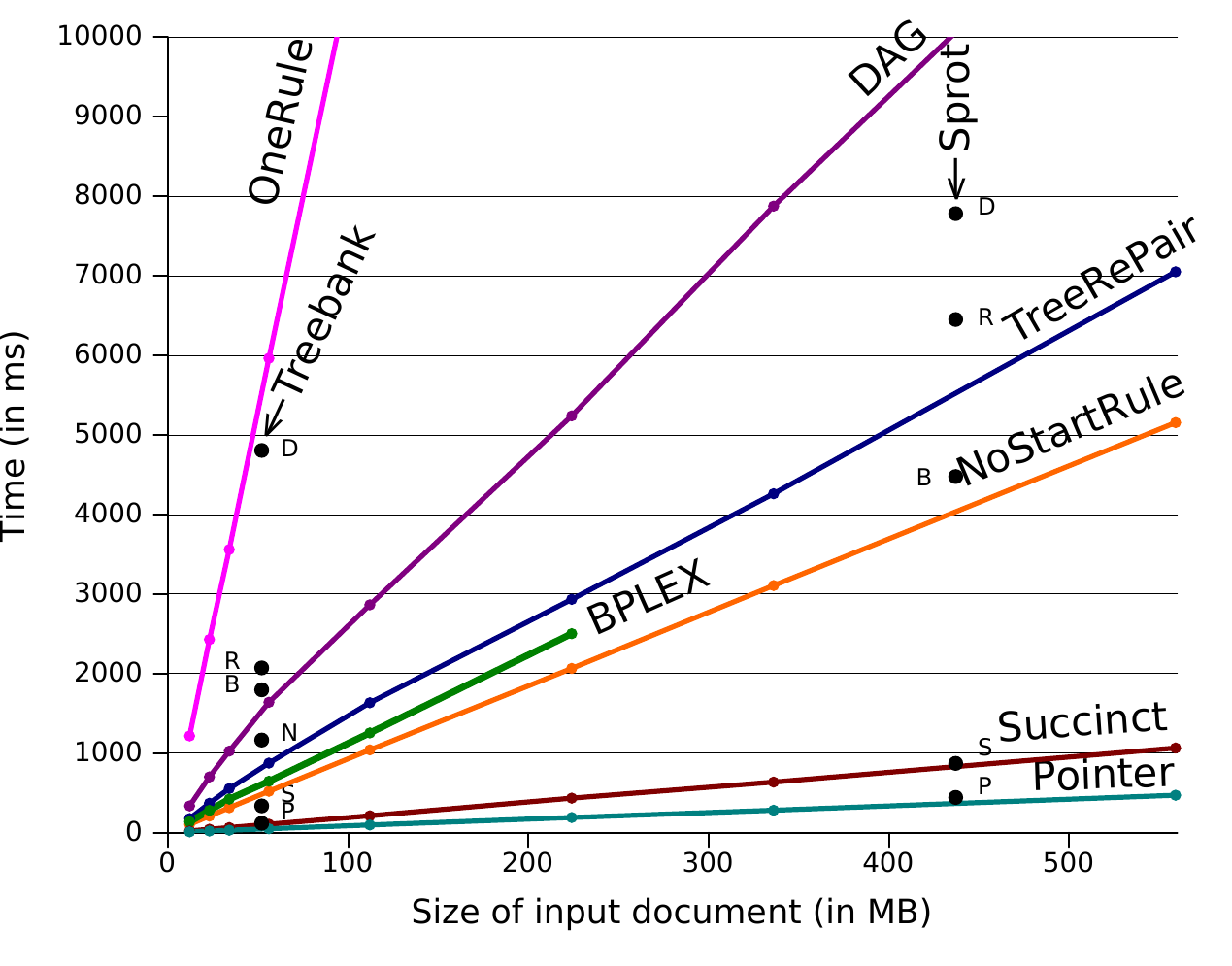}
}
\vspace*{-4mm}
\caption{Iterative tree traversals over XMark}
\label{fig:travIt}
\end{figure}

The speed of our interface is lower-bounded by the speed
of the start rhs representation. 
Since it takes time to get a single pair out of our node ID sequence data
structure, a plain traversal through the (bp)-start rhs is 
slower than a traversal through ``Succinct'' (=the whole XML 
structure tree represented in (bp)).
To see this, we built grammars that have no nonterminals 
(except Start) but store the complete tree in their start rhs.
The full traversal speed of
these grammars is shown as \emph{OneRule} in 
Figures~\ref{fig:travRec} and~\ref{fig:travIt}.
It is also possible to transform the Start rhs into bCNF.
Intuitively, this will introduce as many new nonterminals as there
are nodes in the Start rhs. 
If we apply this to the OneRule grammars of
before, then we obtain \emph{NoStartRule} grammars in which each node is
explicitly represented by a nonterminal.
%(due to bCNF a node would actually use two new nonterminals, 
%but we reduced this to about $1.5$ by ``\_T-sharing'', as explained below). 
The traversal speed
over such grammars should be comparable to that of pointers,
because this is similar to a pointer-based representation.
Again, this is not exactly the case, because of the additional
overhead implied by our node ID data structure. 
%\iffalse
%\fi
Finally, compressed grammars: we
test (binary tree) DAGs, BPLEX, and TreeRePair grammars.
The resulting traversal speeds for iterative
full traversals are shown in Figure~\ref{fig:travIt}.
For recursive traversals the graph looks similar:
all run times are about twice as fast as in the iterative graph,
except for ``Pointer'' which stays the same. The only big 
difference is that for recursive, the DAG line is in between the 
TreeRePair and the NoStartRule lines. Note that for
recursive traversals we added a data structure called ``node pool''
which realizes prefix-sharing of node IDs. Without such
a data structure, recursive traversals are roughly
ten times slower (due to dynamic allocation of node IDs). 
\newcommand{\B}[1]{\textbf{#1}}
\begin{figure}[ht!]
\centerline{
{
\small
\begin{tabular}{|l|*{6}{@{\hspace{3.8pt}}c@{\hspace{3.8pt}}|}}
\hline
XMark & 12MB & 56MB & 112MB & 224MB & 336MB & 559MB \\
\hline
RePair & \B{.1} & \B{.4} & \B{.8} & \B{1} & \B{1.5} & \B{2} \\
BPLEX & .2 & 1 & 3 & 5 & - & - \\
DAG & .6 & 3 & 5 & 9 & 11 & 15 \\
OneRule & 1 & 6 & 12 & 24 & 36 & 60 \\
NoStartRule & 7 & 34 & 68 & 137 & 205 & 342 \\
Succinct & 1 & 3 & 7 & 13 & 19 & 32 \\
Pointer & 13 & 67 & 134 & 267 & 401 & 670 \\
\hline
\end{tabular}
}
}
\caption{Space requirement for iterative traversals}
\label{tab:MemConsumption}
\end{figure}
Through profiling we found
the reason why DAG traversals are
much slower in the iterative case:
DAG grammars we have about eight times more calls to 
the parent function (in the start rhs). The number of these
calls is approximately equal to the number of nodes in
the start rhs. As shown in Figure~\ref{tab:Sizes},
the size of the start rhs is about eight times more 
than those of BPLEX and RePair.
Note that for NoStartRule grammars the number of nonterminals 
equals two times the number of non-\_T-nodes of the document,
plus the number of \_T-nodes of the document.
This is because we use one fixed nonterminal to 
represent \_T-nodes, i.e., we hash-cons all \_T-subtrees.
The OneRule grammars have $(2n-1)$-many nodes in
the start rhs, because for every binary node there is an
additional null-tree.

To summarize the time/space trade-off: 
for recursive traversals,
compared to succinct trees our interface (using TreeRePair grammars)
is $5$--$6$ times slower and uses $3$ times less space,
while it is $12$ times slower and uses $18$ times less space
when compared to pointers.
For iterative traversals we are $7$ times and $15$--$16$ times
slower compared to succinct and pointers, respectively, and use
$9$--$15$ and $167$--$309$ times less space, respectively.

%{\bf Before we go into XPath, show traversal speed for\\
%- chunk-wise\\
%- rule-wise}

\subsection{Counting}

Figure~\ref{fig:xmarkrun} shows timings for XPath counting over
116MB, 1GB, and 11GB XMark files. 
The queries Q01--Q08 and Q13--Q16
are shown in Figure~\ref{fig:treequeries} while 
queries X1--X3 
(taken from~\cite{DBLP:journals/pvldb/GrimsmoBH10})
are shown in Figure~\ref{fig:materialize}.
For TinyT we load our base index plus the jump table. 
For SXSI and MonetDB it was beneficial to load the
entire \emph{original} document: this gave faster counting times
than processing over an XML document representing 
the XML structure tree.
For Qizx the same holds, but, for queries Q05 and X1 the
XML structure tree gave faster times, as indicated by the
footnote in Figure~\ref{fig:xmarkrun}.
The figure shows counting times for our benchmark
queries of Figure~\ref{fig:treequeries}, 
over our three different XMark files.
We were not able to load the 11G XMark document into
SXSI or Qizx. We did succeed to load it in MonetDB, but
times get rather slow from Q03 onwards, due to disk access.
As can be seen, TinyT is faster than all other systems.
Moreover, count times for TinyT scale with respect to the query:
for XMark116M, all queries run in $<$$7$ms. Similarly for the
other documents. This is in stark contrast to all other systems.

The run-time memory of our count evaluator essentially
consists of the index, plus the hash table for
parameter states, plus number counters for each nonterminal.
This adds about 12 Bytes per (state,NT)-pair.
Typically, for an index of 1MB, we have an
additional 2--3MB of run-time memory.
\begin{figure}
\begin{framed}
{\small
\begin{tabular}{l@{\hspace{3pt}}p{72mm}}
Q01 & /site/regions\\
Q02 & /site/closed\_auctions \\
Q03 & /site/regions/europe/item/mailbox/mail/text/keyword \\
Q04 & /site/closed\_auctions/closed\_auction/annotation/description/\\
& \hfill parlist/listitem\\
Q05 & /site/closed\_auctions/closed\_auction/annotation/description/\\
& \hfill parlist/listitem/parlist/listitem/*//keyword\\
Q06 & /site/regions/*/item\\
Q07 & //listitem//keyword\\
Q08 & /site/regions/*/item//keyword \\
%X1 & /site/closed\_auctions/closed\_auction/annotation/description/\\
%&\hfill text/keyword \\
%X2 & //closed auction//keyword\\
%X3 & /site/closed\_auctions/closed\_auction//keyword\\
%X4 & \\
Q13 & //* \\
Q14 & //*//*\\
Q15 & //*//*//*//*\\
Q16 & //*//*//*//*//*//*//*//* \\
\end{tabular}}
\caption{Benchmark queries over XMark}
\label{fig:treequeries}
\end{framed}
\end{figure}

\renewcommand{\thefootnote}{\fnsymbol{footnote}}

\newcommand{\C}[1]{\textbf{#1}}
\newcommand{\I}[1]{\underline{#1}}
\begin{figure*}[ht!]
%\label{fig:count}
\begin{framed}
\centerline{
{\small
%\footnotesize 
\begin{tabular}{|l|*{15}{@{\hspace{5pt}}c@{\hspace{5pt}}|}}
\hline
 & Q01 & Q02 & Q03 & Q04 & Q05 & Q06 & Q07 & Q08 & X1 & X2 & X3 & Q13 & Q14 & Q15 & Q16 \\
\hline
\multicolumn{16}{l}{\rule{0pt}{10pt}XMark116M, counting}\\
\hline
TinyT & \C{.002} & \C{.002} & \C{1} & \C{.7} & \C{1} & \C{1} & \C{5} & \C{2} &
\C{1} & \C{3} & \C{1} & \C{4} & \C{4} & \C{5} & \C{6}\\
SXSI   & 1 & 1 & 14 & 16 & 24 & 12 & 36 & 31 &
22 & 18 & 18 & 
309 & 309 & 313 & 330\\
Monet & 8 & 8 & 24 & 24 & 33 & 16 & 18 & 22 & 23 & 12 & 15 
& 56 & 196 & 476 & 760 \\
Qizx & 1 & 1 & 17\footnotemark[2] & 26 & 33\footnotemark[2] & 6 & 137 & 53 & 
\C{1\footnotemark[2]} & 130 & 28 
& 112 & 3954 & 20457 & 21014\\
\hline
\multicolumn{16}{l}{\rule{0pt}{10pt}XMark116M, serialization}\\
\hline 
TinyT &
\B{141} & \B{46} & \B{19} & \B{26} & \B{9} & \B{168} & \B{66} &
 \B{66} & \B{9} & \B{22} & \B{18}  & \B{2507} & \B{2247} & \B{1773} & \B{312}\\
SXSI & 199 & 67 & 22 & 57 & 33  & 376  & 138 & 85 & 
31 & 44 & 44 &  
6821 & 6267 & 5063 & 1038\\
Monet & 750 & 238 & 30 & 110 & 38 & 757 & 85 & 86 & 59 & 48 & 53
 & 10977 & 9406 & 6095 & 1636 \\
Qizx & 2550 & 839 & 58 & 456 & 74 & 2721 & 257 & 181 & 
56 & 179 & 84 & 
45157 & 44264 & 8181 & 21680\\
\hline
\multicolumn{16}{l}{\rule{0pt}{10pt}XMark1G, counting}\\
\hline 
TinyT & \C{.004} & \C{.004} & \C{5} & \C{3} & \C{4} & \C{3} &
 \C{13} &
 \C{7} & 3 & \B{6} & \B{3}  & \C{17} & \C{20} & \C{24} & \C{31}\\  
SXSI & 2 & 2 & 107 & 149 & 207 & 79 & 665 & 342 &
156 & 146 & 174 &  
4376 & 4371 & 4382 & 4500 \\
Monet & 11 & 11 & 311 & 353 & 399 & 1191 & 2480 & 1238 & 365 & 295 &
332  & 3026 & 4370 & 6973 & 9673 \\
Qizx & 1 & 1 & 89 & 98 & 115 & 32 & 1266 & 412 & 
\B{1\footnotemark[2]} & 1107 & 195 & 
1015 & $++$ & $++$ & $++$ \\
\hline
\multicolumn{16}{l}{\rule{0pt}{10pt}XMark1G, serialization}\\ 
\hline 
TinyT &
\B{1225} & \B{408} & \B{91} & \B{201} & \B{67} & \B{1454} & \B{412} &
 \B{398} & \B{68}  & \B{152} & \B{139} & \B{23987} & \B{21213} & \B{18109} & \B{2339}\\
SXSI & 1922 & 639 & 214 & 597 & 325  & 3598  & 2381 & 1606 & 
304 & 393 & 258 &
73126 & 67169 & 55564 & 12229\\
Monet & 13903 & 4287 & 3484 & 4099 & 4014 & 12429 & 25557 &
 12618 & 4178 & 4043 & 4130 & 115327 & 98622 & $\star$ & 112212 \\
Qizx & 24478 & 7767 & 199 & 203 & 4015 & 24008 & 2432 & 1632 & 
520 & 375 & 562 & 
325794 & $\star\star$ & $\star\star$ &$\star\star$ \\
\hline
\multicolumn{16}{l}{\rule{0pt}{10pt}XMark11G, counting}\\ 
\hline 
TinyT &
\B{.004} & \B{.004} & \B{11} & \B{7} & \B{9} & \B{8} & \B{30} &
 \B{14} & \B{7} & \B{12} & \B{8} & \B{29} & \B{36} & \B{41} & \B{53}\\
Monet & 11 & 11 & 3386 & 4017 & 5214 & 10570 & 24815 &
 11079 & 4111 & 3576 & 3802 
 & 103401 & 397319 & 879057 & $++$ \\
\hline
\end{tabular}
}
}
{\small
\rule{0pt}{10pt}$++$: Running time exceeded 20 minutes.~~~~~~$\star$:
MonetDB server ran out of memory.~~~~~~$\star\star$: Qizx/DB ran out of
memory.
\\[5pt] 
\footnotemark[2]
Time is over the stripped XML structure tree document
of Figure~\ref{fig:structureTree}
(faster than over original document)
%({\bf 13ms}).
%\\[5pt] We mark in \B{bold face} the fastest query execution
%time and we \I{underline} the fastest execution and serialization
%time. 
}
\caption{Times (in ms) for XMark benchmark queries}
\label{fig:xmarkrun}
\end{framed}
\end{figure*}

%\subsubsection*{Comparison to systems for restricted queries}

%\paragraph*{Label Queries}
\subsubsection*{Label Queries}
A label query is of the form //{\it lab\/} and counts the number of
{\it lab\/}-labeled nodes in the document.
Several specialized indexes can be used for fast label-query
execution. Obviously, such queries are not very interesting
(and could be solved through a small extra table).
But, for some systems, such as SXSI, those queries can easily be
restricted to a subtree range. This gives more flexibility; for
instance, count queries such as
//$a$//$b$ could be optimized by moving through the top-most $a$-nodes, and
summing the subtree counts of //$b$ for each such node.

When we write ``SXSI'' in Figures~\ref{fig:lab-queries}
and~\ref{fig:simple_sizes} we 
mean the tree structure index of SXSI. The latter uses
several copies (one per label) of the balanced parenthesis 
structure~\cite{sadnav10}, and compresses those using
sarrays~\cite{okasad07} (uncompressed copies would be even much
larger: 1.8M for one copy of the XMark116M document, times 88 
labels gives 158MB, while with sarrays SXSI only needs 25M).
Intuitively, the private copy of the parenthesis structure for a
given label $\text{\it lab\/}$ indicates only the 
$\text{\it lab\/}$-labeled nodes of the document. Thus, to 
execute the query //category, SXSI accesses the category-copy
of the parenthesis structure and asks for the number of ones
in this structure (realized by the ``rank'' operation which is
efficiently implemented for sarrays).
%We are not aware whether 
%the index of~\cite{ferlucmanmut05,DBLP:conf/www/FerraginaLMM06} can
%also support //lab queries within a given subtree only.
The sizes of the different indexes are shown in
Figure~\ref{fig:simple_sizes}. 

As the timings in Figure~\ref{fig:lab-queries} show,
SXSI is the fastest for such queries, and delivers
constant time.
In the figure ``Fer+'' refers to an implementation 
of~\cite{ferlucmanmut05,DBLP:conf/www/FerraginaLMM06} which
was kindly supplied to us by Francisco Claude
(see the next section for more details).
In this implementation the speed depends on the
selectivity of the query.
\begin{figure}[ht!]
%\begin{framed}
{\small
\centerline{
\begin{tabular}{|l|*{4}{@{\hspace{4pt}}c@{\hspace{4pt}}|}}
\hline
{\small {\bf XMark116M}} &  //category & //price & //keyword & //@*\\
\hline
\#results & 1,040 & 10,140 & 73,070 & 394,611\\
\hline
TinyT & 4.3 & 4.3 & 4.3 & 4.3\\
TinyT+jump & 3.1 & 3.2 & 4.8 & 4.5\\
Fer+ & 0.1 & 0.2 & 1.9 & 10.3\\
SXSI (subtreeTag)& \B{0.02} & \B{0.02} & \B{0.02} & \B{0.02}\\
\hline
\end{tabular}
}
}
\caption{Label queries, counting (in ms)}
\label{fig:lab-queries}
%\end{framed}
\end{figure}

\begin{figure}[ht!]
%\begin{framed}
{\small
\centerline{
\begin{tabular}{|l|*{5}{@{\hspace{7pt}}c@{\hspace{7pt}}|}}
\hline
& TinyT & TinyT+j & Fer+ & Fer-j & SXSI\\
\hline 
XMark-116M & 0.7 & 1.1 & 17.0 & 2.3 & 25.0\\
%\hline
XMark-1G & 2.5 & 5.6 & 163.9 & 22 & 212.9\\
\hline
\end{tabular}
}
}
\caption{Index sizes (in MB)}
\label{fig:simple_sizes}
%\end{framed}
\end{figure}

%\iffalse

%\paragraph*{Simple paths}
\subsubsection*{Simple paths}
An XPath query of the form
$//a_1$/$a_2$/$\cdots$/$a_n$, 
where $a_1,\dots,a_n$ are  label names,
is called \emph{simple path}.
Note that each $a_i$ must be an element name, i.e.,
the wildcard-star (*) is not allowed.
Such queries can be handled by the specialized index of
Ferragina et 
al.~\cite{ferlucmanmut05,DBLP:conf/www/FerraginaLMM06}.
In fact, that index can even materialize result nodes, 
but not by pre-order numbers. We therefore did not include 
it in Section~\ref{sect:Materialization}.
We use our own implementation 
of~\cite{ferlucmanmut05,DBLP:conf/www/FerraginaLMM06},
called ``Fer+''. It is optimized for speed, not for size.
Their own java implementation 
(\url{http://www.di.unipi.it/~ferragin/Libraries/xbwt-demo.zip})
produces
much smaller indexes, see column ``Fer-j'' in 
Figure~\ref{fig:simple_sizes}, but also performs much slower.
For instance, it uses 476ms for query q1 and takes 106ms
for the query q2.
\begin{figure}[ht!]
\begin{framed}
{\small
\centerline{
\begin{tabular}{lcl}
q1 & = & //site/categories/category\\
q2 & = & //open\_auctions/open\_auction/annotation\\
q3 & = & //description/text/text()\\
q4 & = & //text/text()\\[1.5mm]
\end{tabular}
}
%\begin{small}
\centerline{
\begin{tabular}{|l|*{4}{@{\hspace{5pt}}c@{\hspace{5pt}}|}}
\hline
{\bf XMark116M} & q1 & q2 & q3 & q4\\
\hline
\#results & 1,040 & 12,480 & 90,147 & 304,514\\
\hline
TinyT & 4.5 &  4.5 & 4.6 & \B{4.4}\\
TinyT+jump & 3.1 & 3.6 & 5.0 & 5.0\\
Fer+ & \B{0.2} & \B{0.4} & \B{2.6} & 8.4\\
\hline
\multicolumn{4}{l}{\ }\\[-1pt]
\hline
{\bf XMark1G} & q1 & q2 & q3 & q4\\
\hline
\#results & 10,000& 120,000& 868,393 & 2,931,050\\
\hline
TinyT & 21.9 & 22.0 & 22.1 & 19.6 \\
TinyT+jump & 3.0 & 5.8 & \B{14.7} & \B{14.4} \\
Fer+ & \B{0.2} & \B{3.2} & 23.8 & 80.8\\
\hline
\end{tabular}
}
%\end{small}
%\vspace*{1mm}
}
\caption{Simple path queries, counting (in ms)}
\label{fig:simple}
\end{framed}
\end{figure}

Our ``Fer+'' implementation is 
fast for queries with low selectivity and slow for those with 
large selectivity. This can be seen in Figure~\ref{fig:simple}:
for q1 which has the lowest selectivity, Fer+ is 15-times faster
than TinyT, while for q4 TinyT is slightly faster than Fer+.
For larger XMark sizes the relative performance of TinyT is 
better, due to compression: for XMark1G, TinyT is already
faster for q3, and is faster by a factor of $>4.5$ for query q4. 
\iffalse
Note that query q1 is a somewhat  ``artificial'' //-query: it starts
with //site, while XMark documents only have a single site-node (at the
root). If we rewrite q1 into /site/categories/category, then
the running time of TinyT falls from 3ms to 0.06ms (for both
TinyT and TinyT+jump) on XMark116M;
similar but less severe for q2, if its leading //open\_auctions is
changed to /*/open\_auctions then running time drops from 3.8ms to
2.5ms (with TinyT).
\fi

\subsection{Serialization}

Figure~\ref{fig:xmarkrun} shows timings for serialization over
116MB and 1GB XMark files. 
TinyT gave the fastest times for all our serialization
experiments. 
For printing a single subtree (e.g., Q01 and Q02) 
TinyT is about $1.5$-times faster than the next fastest system (SXSI).
For larger result sets the time difference is bigger.
The largest time difference is 
for Q07 (over XMark1G):
TinyT serializes $5.8$-times faster than SXSI. 
For the same query over XMark116M the difference
is only $2.1$. This suggests that the speed-up is related to
the compression in our index.
This is interesting, because one would expect that the 
pure XML serialization time will dominate 
query evaluation and book keeping times. 

For TinyT, 
we load all indexes shown in Figure~\ref{fig:index_breakUp},
together with a ``text collection''. The latter gives access
to getText($i$), the $i$-th value (text or attribute) 
of the document. Our text collection stores all text
(consecutively) in a huge memory buffer.
This takes space (size of the file minus ``Non-Text'' value 
in Figure~\ref{fig:docs}, e.g., 82MB for XMark116M. 
We use a simple data structure
to map from text numbers to begin positions in the buffer.
During serialization we opted for speed, not space.
Recall from Section~\ref{sect:chunk} our serialization process:
we first build tag sequences of all document subtrees 
to be output, together with copy instructions that
point into those sequences. These tag sequences still contain
\_T tags. After evaluation, XML serialization starts by
(1) writing full XML subtrees by correctly replacing \_T nodes
by their text values and also (2) replacing copy instructions by
their correct serialization.
%We start serialization \emph{during} evaluation, because
%already then we generate the result tag sequences (and copy
%instructions). But actual XML serialization takes place
%afterwards in a separate step. 

\subsection{Materialization}
\label{sect:Materialization}

Initially TinyT was built for fast evaluation of 
XPath count queries. 
Later we realized its usefulness for fast serialization; the key idea
was to avoid materialization of result nodes and to print
directly in parallel with query evaluation. 
The running times for both counting and printing are highly
competitive, as can be seen in Figure~\ref{fig:xmarkrun}.
We also wanted to compare to specialized systems such as
implementations of twig queries. Twig queries
have been studied extensively both from
a theoretical and an implementational view point.
They belong to the most highly optimized XPath queries 
(see, e.g.,~\cite{DBLP:journals/pvldb/GrimsmoBH10,DBLP:conf/dbkda/GrimsmoBT10})
and the references those articles) 
Twig query implementations materialize several context
nodes per query result. This is different from XPath semantics in which
one node only is selected at a time.
Clearly it would not be fair
to compare our count evaluator with a twig implementation that
materializes (even multiple nodes per result).
We decided to build a materializer for TinyT which
produces pre-order numbers of the result nodes. 
This was done in short time, by essentially reusing the
code of the serializer, and indeed, doing a fair amount
of serialization in memory during materialization. 
Certainly, this implementation is far from optimal; it would
be much more efficient to work over node offset numbers, 
rather than serialized XML tag sequences. 
As the experiments in Figure~\ref{fig:materialize} show, TinyT is the
fastest only for query X3, while for X1 and X2 
XLeaf and TJStrictPre are the fastest, respectively.
We believe that a more efficient implementation of materializing
over TinyT can be considerable faster, 
ca. 2--3 times slower than counting.
%While then TinyT would be faster than the two twig implementations
%for query X2, it would still be slower than XLeaf for X1:
%XLeaf only needs only $3.6$ms (while counting is $3$ms with TinyT).

\begin{figure}[ht!]
\begin{framed}
{
\small
\centerline{
\begin{tabular}{lcl}
X1 & = &
/site/closed\_auctions/closed\_auction/annotation/\\
& & \hfill description/text/keyword\\
X2 & = & //closed auction//keyword\\
X3 & = & /site/closed\_auctions/closed\_auction//keyword\\[1.5mm]
\end{tabular}
}
\centerline{
\begin{tabular}{|l|*{3}{@{\hspace{8pt}}c@{\hspace{8pt}}|}}
\hline
{\bf XMark1G} & X1 & X2 & X3\\
\hline
\#results & 40,726 & 124,843& 124,843\\
\hline
TinyT & 25 & 52 & 34\\
TinyT+jump &  36 & 38 & \B{33}  \\
SXSI & 183 & 174 & 164\\
MonetDB & 235 & 180 & 199\\
TJStrictPre~\cite{DBLP:journals/pvldb/GrimsmoBH10} & 145 & \B{21} & 40\\
XLeaf~\cite{DBLP:conf/dbkda/GrimsmoBT10} & \B{3.6} & 47 & 47\\
\hline
\end{tabular}
}
%\end{small}
%\vspace*{1mm}
}
\caption{Twig queries, materialization (in ms)}
\label{fig:materialize}
\end{framed}
\end{figure}

\subsection{Compression Behavior}

Our algorithms that execute without decompression
directly on the grammar
such as rule-wise XPath counting or chunk-wise XPath
serialization, both do one pass through the grammar.
Thus, the running time of these algorithms is 
strongly influenced by the size of the grammar.
TreeRePair generates smaller grammars than
BPLEX (about half the size, in terms of numbers of edges)~\cite{lohmanmen10},
which itself makes smaller grammars than DAGs~\cite{buslohman08}.
Therefore, our count and serialize XPath evaluators 
run fastest over grammars produced by TreeRePair.
The size of the start rhs is important too, because
access is slower and more complicated than over the recursive rules
(compare run times of OneRule with NoStartRule in 
Figures~\ref{fig:travRec} and~\ref{fig:travIt}).
BPLEX grammars have relatively small start rhs's, 
but, the problem with those grammars is the high rank of
nonterminals:
often $10$ or more parameters are needed in order
to get small grammars. 
Figure~\ref{tab:Sizes} shows information about the grammars
used in Section~\ref{sect:traversal} for the recursive and
iterative traversals.
The underlined number is the size of the start rhs,
and the numbers below are: number of nonterminals
of rank $0$, number nonterminals of rank $1$, etc.
After transformation to bCNF, DAGs have one parameter.
TreeRePair was instructed to produce grammars of rank $\leq 1$
(which gives grammars of rank 2 in bCNF). 
This gave the best performance for our traversal experiments.
Note that for the XPath evaluators, TreeRePair grammars of
rank $2$ are optimal (which have rank $3$ in bCNF).
For BPLEX we generated
grammars that have $12$ parameters in their final bCNF form
(only the first $5$ numbers of nonterminals are shown in the figure).
To see the impact of the size of the grammar, consider
the query //listitem//keyword 
(which is adequate because no skipping takes place in the
start rhs) over XMark116M:
our count evaluator takes 5ms over a TreeRePair (rank $2$) grammar.
In contrast, evaluating over a DAG grammar takes
$28$ms (the time for BPLEX grammars is in the middle: 17ms). 
This is due to the large number of parameters of BPLEX grammars:
If the rank of a grammar is high, then hashing and handling
of parameter states of a nonterminal becomes more expensive
in the automaton evaluation functions of our XPath evaluators.
This can be seen best on two grammars
produced by TreeRePair for XMark11G.
Both grammars are of similar size, but one has rank $8$ while
the other has rank $3$. Our count evaluator takes 38ms
for the first grammar and only 28ms for the second.
Thus, the number of parameters has a large impact.
In summary, setting the maximal rank to $2$ in TreeRePair
(and thus obtaining grammars of rank $3$ in bCNF form),
gives the best trade-off for our evaluators across all
tested documents.
\begin{figure}
{
\small
\centerline{
\begin{tabular}{|l|*{3}{@{\hspace{-3pt}}c@{\hspace{-3pt}}|}}
\hline
{\bf XMark} & 12MB & 116MB & 224MB \\
\hline
RePair & \begin{tabular}{c}\underline{36600}\\828, 1413, 241\\\end{tabular} &
\begin{tabular}{c}\underline{153366} \\22849, 22061, 11511 \\\end{tabular} &
\begin{tabular}{c}\underline{190628} \\45272, 44974, 23889 \\\end{tabular}\\
\hline
BPLEX & 
\begin{tabular}{c}\underline{17704} \\12614, 7113, 411
%\\82, 35, 55, 28, 23
%\\59, 119, 37, 2
\\ 82, 35, \dots
\\\end{tabular} & 
\begin{tabular}{c}\underline{105801}
\\194032, 97515, 613
%\\228, 143, 90, 91
%\\250, 807, 376, 40, 2 
\\ 228, 143, \dots
\end{tabular}
& 
\begin{tabular}{c}\underline{175135} \\384757, 192029, 628
%\\286, 201, 126, 92
%\\386, 1245, 284, 50
\\ 286, 201, \dots
\end{tabular} \\
\hline
DAG & \begin{tabular}{c}\underline{189683} \\12614, 11100 \\\end{tabular} &
\begin{tabular}{c}\underline{894695} \\194032, 172094 \\\end{tabular} &
\begin{tabular}{c}\underline{1168011} \\384757, 340246 \\\end{tabular}\\
\hline
\end{tabular}
}
}
\caption{Size start rhs and number rank-$k$ nonterminals}
\label{tab:Sizes}
\end{figure}

\section{Discussion}

We presented a new structural index for XML and evaluated its
performance for XPath evaluation.
The index is based on a grammar compressed representation
of the XML structure tree.
For common XML documents the corresponding indexes are minuscule.
When executing arbitrary tree algorithms over the index,
a good time-space trade-off is obtained.
For certain simple XPath tasks such as result node counting,
impressive speed-ups can be achieved. 
Our rudimentary XPath implementation over this index outperforms
the fastest known systems (MonetDB and Qizx), both for counting
and for serialization. We built and experimental materializer which
is competitive with the state-of-the art twig query implementations. 
We believe that our system is useful for other
XPath evaluators and XML databases.
It can be used for selectivity computation of structural queries,
and for fast serialization. It will be interesting to extend
our current XPath evaluators to handle filters, and also to
handle data value comparisons. For the latter 
bottom-up evaluator as in~\cite{DBLP:journals/pvldb/ManethN10} could 
be built,
which first searches over the text value store, and then 
verifies paths in the tree, in a bottom-up way. For such queries
the SXSI system~\cite{arr+10} is highly efficient.
We do not expect to achieve faster run times with our index,
but think that run times similar to those of SXSI can be achieved.
This is a large improvement, because the space requirement of our index 
is much smaller than that of SXSI. 

It would be interesting to add specialized indexes which allow 
more efficient running times for simple queries, such as 
simple path queries of the form $//a_1/a_2/\dots/a_m$.
Over strings, the self index of~\cite{clanav10} allows to find occurrences
of such queries in time logarithmic in the number of rules of the grammar.
Can their result be generalize to the tree case?
In terms of extraction (decompression) there are new results 
for DAGs~\cite{bilae11}
that run in time logarithmic in the number of edges
of the DAG. Can this result be generalized from DAGs to our
SLT grammar?

\subsection*{Acknowledgments}

We are thankful to Francisco Claude for providing to us an
alternative implementation 
of~\cite{ferlucmanmut05,DBLP:conf/www/FerraginaLMM06}.
Many thanks to the authors 
of~\cite{DBLP:journals/pvldb/GrimsmoBH10,DBLP:conf/dbkda/GrimsmoBT10}
and to Nils Grimsmo in particular, for providing to us his implementations
used for those articles.
Finally, we thank Xavier Franc from XMLmind for providing to us
their implementation of Qizx, and for assisting us in testing
against their system.

\bibliographystyle{abbrv}
\bibliography{mybib.bib}

\begin{thebibliography}{10}

\bibitem{arr+10}
D.~Arroyuelo, F.~Claude, S.~Maneth, V.~M{\"a}kinen, G.~Navarro, K.~Nguyen,
  J.~Sir{\'e}n, and N.~V{\"a}lim{\"a}ki.
\newblock Fast in-memory {XPath} search using compressed indexes.
\newblock In {\em ICDE}, pages 417--428, 2010.

\bibitem{DBLP:journals/jacm/Baeza-YatesG96}
R.~A. Baeza-Yates and G.~H. Gonnet.
\newblock Fast text searching for regular expressions or automaton searching on
  tries.
\newblock {\em J. ACM}, 43(6):915--936, 1996.

\bibitem{bilae11}
P.~Bille, G.~M. Landau, R.~Raman, K.~Sadakane, S.~R. Satti, and O.~Weimann.
\newblock Random access to grammar-compressed strings.
\newblock In {\em SODA}, 2011.
\newblock To appear.

\bibitem{DBLP:conf/sigmod/BonczGKMRT06}
P.~A. Boncz, T.~Grust, M.~van Keulen, S.~Manegold, J.~Rittinger, and
  J.~Teubner.
\newblock {MonetDB/XQuery}: a fast {XQuery} processor powered by a relational
  engine.
\newblock In {\em SIGMOD Conference}, pages 479--490, 2006.

\bibitem{bungrokoc03_short}
P.~Buneman, M.~Grohe, and C.~Koch.
\newblock Path queries on compressed {XML}.
\newblock In {\em VLDB}, pages 141--152, 2003.

\bibitem{buslohman08}
G.~Busatto, M.~Lohrey, and S.~Maneth.
\newblock Efficient memory representation of {XML} document trees.
\newblock {\em Inf. Syst.}, 33:456--474, 2008.

\bibitem{clanav10}
F.~Claude and G.~Navarro.
\newblock Self-indexed grammar-based compression.
\newblock To appear in Fundamenta Informaticae, 2010.

\bibitem{DBLP:books/mg/CormenLRS01}
T.~H. Cormen, C.~E. Leiserson, R.~L. Rivest, and C.~Stein.
\newblock {\em Introduction to Algorithms, Second Edition}.
\newblock The MIT Press and McGraw-Hill Book Company, 2001.

\bibitem{ferlucmanmut05}
P.~Ferragina, F.~Luccio, G.~Manzini, and S.~Muthukrishnan.
\newblock Structuring labeled trees for optimal succinctness, and beyond.
\newblock In {\em FOCS}, pages 184--196, 2005.

\bibitem{DBLP:conf/www/FerraginaLMM06}
P.~Ferragina, F.~Luccio, G.~Manzini, and S.~Muthukrishnan.
\newblock Compressing and searching {XML} data via two zips.
\newblock In {\em WWW}, pages 751--760, 2006.

\bibitem{fisman07}
D.~K. Fisher and S.~Maneth.
\newblock Structural selectivity estimation for {XML} documents.
\newblock In {\em ICDE}, pages 626--635, 2007.

\bibitem{DBLP:journals/is/FletcherGWGBP09}
G.~H.~L. Fletcher, D.~V. Gucht, Y.~Wu, M.~Gyssens, S.~Brenes, and J.~Paredaens.
\newblock A methodology for coupling fragments of {XPath} with structural
  indexes for {XML} documents.
\newblock {\em Inf. Syst.}, 34:657--670, 2009.

\bibitem{DBLP:conf/vldb/GoldmanW97}
R.~Goldman and J.~Widom.
\newblock Dataguides: Enabling query formulation and optimization in
  semistructured databases.
\newblock In {\em VLDB}, pages 436--445, 1997.

\bibitem{DBLP:journals/pvldb/GrimsmoBH10}
N.~Grimsmo, T.~A. Bj{\o}rklund, and M.~L. Hetland.
\newblock Fast optimal twig joins.
\newblock {\em PVLDB}, 3(1):894--905, 2010.

\bibitem{DBLP:conf/dbkda/GrimsmoBT10}
N.~Grimsmo, T.~A. Bj{\o}rklund, and {\O}.~Torbj{\o}rnsen.
\newblock Xleaf: Twig evaluation with skipping loop joins and virtual nodes.
\newblock In {\em DBKDA}, pages 204--213, 2010.

\bibitem{DBLP:conf/icde/HeY04}
H.~He and J.~Yang.
\newblock Multiresolution indexing of {XML} for frequent queries.
\newblock In {\em ICDE}, pages 683--694, 2004.

\bibitem{DBLP:conf/sigmod/KaushikBNK02}
R.~Kaushik, P.~Bohannon, J.~F. Naughton, and H.~F. Korth.
\newblock Covering indexes for branching path queries.
\newblock In {\em SIGMOD Conference}, pages 133--144, 2002.

\bibitem{DBLP:conf/icde/KaushikSBG02}
R.~Kaushik, P.~Shenoy, P.~Bohannon, and E.~Gudes.
\newblock Exploiting local similarity for indexing paths in graph-structured
  data.
\newblock In {\em ICDE}, pages 129--140, 2002.

\bibitem{lam90}
J.~Lamping.
\newblock An algorithm for optimal lambda calculus reductions.
\newblock In {\em POPL}, pages 16--30, 1990.

\bibitem{lohman06}
M.~Lohrey and S.~Maneth.
\newblock The complexity of tree automata and {XPath} on grammar-compressed
  trees.
\newblock {\em Theor. Comput. Sci.}, 363(2):196--210, 2006.

\bibitem{lohmanmen10}
M.~Lohrey, S.~Maneth, and R.~Mennicke.
\newblock Tree structure compression with {RePair}.
\newblock {\em CoRR}, abs/1007.5406, 2010.
\newblock Short version to appaer as paper in {\it DCC'2011}.

\bibitem{lohmansch09}
M.~Lohrey, S.~Maneth, and M.~Schmidt-Schau{\ss}.
\newblock Parameter reduction in grammar-compressed trees.
\newblock In {\em FOSSACS}, pages 212--226, 2009.

\bibitem{DBLP:conf/dexaw/ManethMS08}
S.~Maneth, N.~Mihaylov, and S.~Sakr.
\newblock {XML} tree structure compression.
\newblock In {\em DEXA Workshops}, pages 243--247, 2008.

\bibitem{DBLP:journals/pvldb/ManethN10}
S.~Maneth and K.~Nguyen.
\newblock {XPath} whole query optimization.
\newblock {\em PVLDB}, 3(1):882--893, 2010.

\bibitem{DBLP:conf/icdt/MiloS99}
T.~Milo and D.~Suciu.
\newblock Index structures for path expressions.
\newblock In {\em ICDT}, pages 277--295, 1999.

\bibitem{okasad07}
D.~Okanohara and K.~Sadakane.
\newblock Practical entropy-compressed rank/select dictionary.
\newblock In {\em ALENEX}, 2007.

\bibitem{DBLP:conf/sac/PettovelloF06}
P.~M. Pettovello and F.~Fotouhi.
\newblock {MTree}: an {XML} {XPath} graph index.
\newblock In {\em SAC}, pages 474--481, 2006.

\bibitem{DBLP:conf/sigmod/QunLO03}
C.~Qun, A.~Lim, and K.~W. Ong.
\newblock D(k)-index: An adaptive structural summary for graph-structured data.
\newblock In {\em SIGMOD Conference}, pages 134--144, 2003.

\bibitem{DBLP:conf/xsym/RunapongsaPBP04}
K.~Runapongsa, J.~M. Patel, R.~Bordawekar, and S.~Padmanabhan.
\newblock {XIST}: An {XML} index selection tool.
\newblock In {\em XSym}, pages 219--234, 2004.

\bibitem{ryt04_short}
W.~Rytter.
\newblock Grammar compression, {LZ}-encodings, and string algorithms with
  implicit input.
\newblock In {\em ICALP}, pages 15--27, 2004.

\bibitem{sadnav10}
K.~Sadakane and G.~Navarro.
\newblock Fully-functional succinct trees.
\newblock In {\em SODA}, pages 134--149, 2010.

\bibitem{qizx}
{XMLmind Products}.
\newblock Qizx, a fast {XML} database engine fully supporting {XQuery}.
\newblock http://www.xmlmind.com/qizx/qizx.html.

\bibitem{DBLP:conf/sigmod/YiHSY04}
K.~Yi, H.~He, I.~Stanoi, and J.~Yang.
\newblock Incremental maintenance of {XML} structural indexes.
\newblock In {\em SIGMOD Conference}, pages 491--502, 2004.

\end{thebibliography}

\end{document}